# The Law of DeepSeek

*Jyh-An Lee* * *& Xuan Sun* **

ABSTRACT

*Amid the intensifying competition in artificial intelligence between the United States and China, the emergence of the DeepSeek-R1 model has sent significant ripples through the technology sector, capital markets, and policy circles. This Article offers a comprehensive analysis of the legal and policy landscape surrounding DeepSeek, drawing upon its key technical features—including reinforcement learning, mixture-of-experts architecture, multi-head latent attention mechanism, knowledge distillation, and open-source approach.*

*Firstly, DeepSeek's success raises critical questions about the efficacy of the United States' increasingly robust export control measures on chips and semiconductors, components essential for training large language models. Secondly, akin to Chinese technology giants such as TikTok and Huawei, DeepSeek is confronted with information security scrutiny in the United States and other jurisdictions. Thirdly, despite its domain-specific capabilities and competitive API pricing, DeepSeek has faced criticism for producing outputs laden with political and ideological biases, igniting debates over free speech and censorship.*

*Furthermore, this Article delves into intellectual property and contractual concerns stemming from the knowledge distillation technique employed by DeepSeek. Ultimately, it concludes that geopolitical considerations will continue to exert a profound influence on the legal challenges and prospective solutions related to the DeepSeek models.*

---

* Professor and Executive Director, Centre for Legal Innovation and Digital Society (CLINDS), The Chinese University of Hong Kong Faculty of Law.

** Ph.D. Candidate, The Chinese University of Hong Kong Faculty of Law.

This paper was presented at the 2025 GILJ International Law Symposium on "BRICS Plus and the Evolving Technological and Financial World Order" at Gonzaga Law School, as well as at the 25th CLINDS Law & Digital Society Seminar at the Chinese University of Hong Kong Faculty of Law. The authors express their gratitude to Upendra Dev Acharya, Chien-Yi Chang, Dongmin Chen, Farzan Farnia, Taorui Guan, Yangzi Li, Jingwen Liu, Eliza Mik, Agnieszka McPeak, Jue Wang, Chenxin Yang, Qian Yin, Dongsheng Zang, and Peng Zhou for their invaluable discussions and comments. This study was supported by the RGC Humanities and Social Sciences Prestigious Fellowship Scheme (HSSPFS) 2025/26 (Project No.: CUHK 3400625).

## I. INTRODUCTION

Artificial intelligence ("AI") has not only revolutionized human society and the global economy,[1] but has also reshaped the international balance of power. The current "AI arms race" is frequently compared to the space race of the 1960s and the nuclear arms race from the 1950s to the 1980s.[2] Accordingly, it is hardly surprising that some assert that the leading force in AI technology may ultimately determine global leadership.[3] Although AI breakthroughs are occurring worldwide, many contend that the competition ultimately centers on the United States and China.[4]

While it has long been believed that the United States enjoys a commanding lead in AI development, the emergence of the Chinese AI company DeepSeek and its large-language model ("LLM"), DeepSeek-R1, demonstrates that China's progress in AI is swifter than many had assumed,[5] despite U.S. efforts to restrict China's technological advancement.[6] DeepSeek's extraordinary success has sparked concerns in the United States that its most advanced AI products may no longer be able to compete with more affordable Chinese alternatives.[7] Indeed, the U.S. President Donald Trump described DeepSeek as a "wake-up call" for the American technology industry.[8] Former FTC Chair Lina Khan has argued that robust antitrust enforcement against DeepSeek and other Chinese tech firms is critical to national security, warning that insufficient competition leaves the U.S. tech industry vulnerable to Chinese rivals and undermines America's geopolitical influence in the 21st Century.[9]

---

[1] *See, e.g.*, Nicholas Caputo, *"Quiet" Enjoyment: Uncovering the Hidden History of the Right to Attention in Private and Public*, 28 STAN. TECH. L. REV. 120, 129–30 (2025) (noting that large language models, such as ChatGPT27 and DeepSeek's R1, exemplify a new surge of technological innovation poised to transform human society and fundamentally reshape the ways we live and work).

[2] LEE KAI-FU, AI SUPERPOWERS: CHINA, SILICON VALLEY, AND THE NEW WORLD ORDER 94–96, 227–28 (2018).

[3] Radina Gigova, *Who Vladimir Putin Thinks Will Rule the World*, CNN (Sep. 2, 2017, at 01:07 ET), https://edition.cnn.com/2017/09/01/world/putin-artificial-intelligence-will-rule-world/index.html [https://perma.cc/TC6F-PQ4Y]; Mark Minevich, *Whoever Leads In AI Compute Will Lead The World*, FORBES (Jan. 17, 2025, at 01:28 ET), https://www.forbes.com/sites/markminevich/2025/01/17/whoever-leads-in-ai-compute-will-lead-the-world [https://perma.cc/D2UM-Y2U2]; Indermit Gill, *Whoever leads in artificial intelligence in 2030 will rule the world until 2100*, BROOKINGS INST. (Jan. 17, 2020), https://www.brookings.edu/articles/whoever-leads-in-artificial-intelligence-in-2030-will-rule-the-world-until-2100 [https://perma.cc/49UU-KE99].

[4] *See, e.g.*, Neil Savage, *The Race to the Top Among the World's Leaders in Artificial Intelligence*, NATURE INDEX (Dec. 9, 2020), https://www.nature.com/articles/d41586-020-03409-8 [https://perma.cc/4SPD-34PC].

On January 20, 2025, DeepSeek officially launched its DeepSeek-R1 model.[10] Despite preparations underway in China for the year's largest celebration—the Spring Festival—the DeepSeek-R1 model quickly went viral on the Chinese internet, soon drawing global attention.[11] Influential tech investor Marc Andreessen aptly characterized this day as the "AI's Sputnik moment," and praised DeepSeek-R1 model as "one of the most amazing and impressive breakthroughs" he had ever witnessed.[12] Boasting distinct advantages in training cost and performance comparable to proprietary models like ChatGPT, DeepSeek-R1 became the most-downloaded free app on Apple's app store after its U.S. launch.[13] DeepSeek's founder and CEO Liang Wenfeng has been named among *Nature*'s top ten "people who shaped science in 2025," with the prestigious scientific journal describing the forty-year-old entrepreneur as a "Chinese finance whizz" whose breakthrough AI models have "stunned the world."[14]

---

[5] *See, e.g.*, Amay Hattangadi & Samson Hung, *China's DeepSeek Moment*, MORGAN STANLEY (Mar. 18, 2025), https://www.morganstanley.com/im/publication/insights/articles/article_chinasdeepseekmoment_a4.pdf [https://perma.cc/9G4Z-VCZP] (viewing DeepSeek as a signal of China's broader resurgence in innovation and global competitiveness).

[6] *See, e.g.*, Andrew R. Chow, *Why DeepSeek Is Sparking Debates Over National Security, Just Like TikTok*, TIME (Jan. 29, 2025, at 05:44 CT), https://time.com/7210875/deepseek-national-security-threat-tiktok [https://perma.cc/V3US-MZTQ] (arguing that "DeepSeek shows that China's AI has developed much faster than many had believed, despite efforts from American policymakers to slow its progress."); Matt Sheehan & Scott Singer, *What DeepSeek Revealed About the Future of U.S.-China Competition*, FOREIGN POL'Y (Feb. 3, 2025, at 03:01 CT), https://foreignpolicy.com/2025/02/03/deepseek-china-ai-artificial-intelligence-united-states-tech-competition [https://perma.cc/X8V4-FHNN].

[7] *See, e.g.*, Sheehan & Singer, *supra* note 6; *see also* Mihailis E. Diamantis, *Reasonable AI: A Negligence Standard*, 78 VAND. L. REV. 573, 579 n.34 (2025) (describing DeepSeek as a Chinese chatbot that rivals the U.S. leader ChatGPT while operating on less advanced processors and indicating that the emergence of DeepSeek underscores the urgent economic and national security implications of the global AI race).

[8] Kelly Ng et al., *DeepSeek: The Chinese AI App That Has the World Talking*, BBC (Feb. 4, 2025), https://www.bbc.com/news/articles/c5yv5976z9po [https://perma.cc/7NWV-54LJ].

[9] Lina M. Khan, *Stop Worshiping the American Tech Giants*, N.Y. TIMES (Feb. 4, 2025), https://www.nytimes.com/2025/02/04/opinion/deepseek-ai-big-tech.html [https://perma.cc/V3US-MZTQ].

[10] *DeepSeek-R1 Release*, DEEPSEEK (Jan. 20, 2025), https://api-docs.deepseek.com/news/news250120 [https://perma.cc/NW56-3JQF].

[11] Brenda Goh & Eduardo Baptista, *Chinese AI Start-up DeepSeek Pushes US Rivals with R1 Model Upgrade*, REUTERS (May 29, 2025, at 10:32 CT), https://www.reuters.com/world/china/chinas-deepseek-releases-an-update-its-r1-reasoning-model-2025-05-29 [https://perma.cc/G2FP-ZX3T].

[12] David Goldman, *What Is DeepSeek, the Chinese AI Startup That Shook the Tech World?*, CNN (Jan. 27, 2025, at 16:55 ET), https://www.cnn.com/2025/01/27/tech/deepseek-ai-explainer [https://perma.cc/3FXE-MS9L].

In China, the DeepSeek model was swiftly adopted by leading technology giants. For example, on February 8, 2025, Tencent Cloud launched a DeepSeek Application Programming Interface ("API") featuring advanced internet search capabilities,[15] while AliCloud integrated the comprehensive version of the DeepSeek API into its services.[16] Home appliance manufacturers such as Midea Group and Haier have incorporated DeepSeek technology into their televisions and refrigerators.[17] Simultaneously, government agencies at various administrative levels across the country embraced the DeepSeek model as a powerful tool to enhance operational efficiency.[18] Regions such as Guangdong, Jiangsu, Inner Mongolia, and Jiangxi have announced the integration of DeepSeek into their local online public service platforms.[19] State-owned enterprises, including China Sinopec and China Unicom, have likewise deployed the DeepSeek model to drive innovation.[20]

In the Western world, while governments maintain a cautious stance toward the DeepSeek model,[21] prominent technology companies have been quick to recognize its potential and incorporate it into their products.[22] Immediately after DeepSeek's launch, DeepSeek-R1 was made available via Microsoft's Azure AI.[23] NVIDIA and Amazon.com also provide users with access to DeepSeek's models.[24] In addition, developers can now easily build specialized agents using DeepSeek-R1 as an NVIDIA NIM microservice on build.nvidia.com.[25] Notably, in April 2025, BMW announced a collaboration with DeepSeek to advance the intelligence of its in-car AI systems.[26] Further underscoring DeepSeek's global push, it plans to launch the first country-specific version of its namesake chatbot in Italy and has pledged to curb model "hallucinations" to meet the country's regulatory standards.[27]

---

[13] Ng et al., *supra* note 8; *see also* Bryan Druzin, Anatole Boute & Michael Ramsden, *Confronting Catastrophic Risk: The International Obligation to Regulate Artificial Intelligence*, 46 MICH. J. INT'L L. 173, 178 (2025) (highlighting the emergence of Chinese LLMs such as DeepSeek and Alibaba's Qwen2.5-Max and the growing geopolitical dimensions of the U.S.-China AI competition).

[14] Elizabeth Gibney, *The Chinese Finance Whizz Whose DeepSeek AI Model Stunned the World*, NATURE (Dec. 8, 2025), https://www.nature.com/articles/d41586-025-03845-4 [https://perma.cc/4PNG-2VG2].

[15] 腾讯云 [Tencent Cloud], *率先支持！腾讯云上线 DeepSeek API 接口+联网搜索* [*Tencent Cloud Launches DeepSeek API and Internet Search Integration*], TENCENT CLOUD DEVELOPER CMTY. (Feb. 8, 2025, at 04:00 CT), https://cloud.tencent.com/developer/article/2494477 [https://perma.cc/L8GV-7AN2]. Tencent Cloud is a major cloud computing platform from the Chinese tech giant Tencent. *提供全球领先的云计算服务* [*Providing Globally Leading Cloud Computing Services*], 腾讯云 [TENCENT CLOUD], https://cloud.tencent.com/about?Is=sdk-topnav [https://perma.cc/Y9WU-5MYV].

[16] 阿里云 [Alibaba Cloud], 一文轻松拿下 DeepSeek 满血版的 Api 接口接入 [*Quick Access to the Full-Capability DeepSeek API via the Alibaba Cloud Bailian Platform*], ALIBABA CLOUD DEVELOPER CMTY. (Mar. 4, 2025), https://developer.aliyun.com/article/1655480. Alibaba Cloud is one of the Chinese leading cloud computing and artificial intelligence technology companies under the Alibaba Group. *了解阿里云* [*About Alibaba Cloud*], 阿里云 [ALIBABA CLOUD], https://cn.aliyun.com/about?from_alibabacloud=&spm=5176.28507329.J_4VYgf18xNlTAyFFbOuOQe.d_menu_9.4c982868mDlF06 [https://perma.cc/93N2-RLRG ] (last visited Mar. 1, 2026).

[17] Ben Jiang, Coco Feng & Wency Chen, *How China's Open-Source AI is Helping DeepSeek, Alibaba Take on Silicon Valley*, S. CHINA MORNING POST (July 19, 2025, at 10:00 CT), https://www.scmp.com/tech/big-tech/article/3318747/how-chinas-open-source-ai-helping-deepseek-alibaba-take-silicon-valley [https://perma.cc/8TH9-3897].

[18] 环球网 [Huanqiu], *多地政务系统、中企巨头，接入 DeepSeek!* [*Multiple Government Portals and Major Domestic Firms Integrate DeepSeek!*], GLOBAL TIMES (Feb. 17, 2025, at 00:55 CT), https://china.huanqiu.com/article/4LWVHFjBBQM [https://perma.cc/3Z6J-UC7T].

[19] *Id.*

[20] *Id.*

What has truly unsettled the American AI sector is that DeepSeek has revealed an alternative model development strategy, challenging several long-held assumptions in the technical community, such as the scaling law,[28] the pivotal role of high-end semiconductors,[29] and the perceived supremacy of proprietary models over open-source models.[30] It is reported that training DeepSeek's base model cost merely $5.6 million,[31] compared to the over $100 million spent on OpenAI's GPT-4.[32] While DeepSeek's peers reportedly rely on more than 16,000 chips to train their models, DeepSeek's engineers used only approximately 2,000 NVIDIA chips.[33] Remarkably, in domains such as mathematics and coding, DeepSeek demonstrates exceptional performance,[34] often equaling or even surpassing the capacity of its Western counterparts.[35] Commentators suggested that DeepSeek's success signaled a bifurcated future for global AI development: one path characterized by the capital-intensive, semiconductor-dependent approach of the United States; the other, represented by China's efficiency-driven model.

---

[21] *See infra* text accompanying notes 165–75.

[22] *Western companies are experimenting with DeepSeek*, The ECONOMIST (Mar. 13, 2025), https://www.economist.com/business/2025/03/13/western-companies-are-experimenting-with-deepseek [https://perma.cc/7RZY-DFTP].

[23] Asha Sharma, *DeepSeek R1 is Now Available on Azure AI Foundry and GitHub*, MICROSOFT AZURE BLOG (Jan. 29, 2025), https://azure.microsoft.com/en-us/blog/deepseek-r1-is-now-available-on-azure-ai-foundry-and-github [https://perma.cc/JP6T-4YX9].

[24] Jiang, Feng & Chen, *supra* note 17.

[25] Erik Pounds, *DeepSeek-R1 Now Live With NVIDIA NIM*, NVIDIA BLOG (Jan. 30, 2025), https://blogs.nvidia.com/blog/deepseek-r1-nim-microservice [https://perma.cc/NF9X-UDX6]. This DeepSeek-R1 NIM microservice simplifies deployments by using support for industry-standard APIs. By using this service, companies can promote security and data privacy on their preferred computing infrastructure.

[26] Horatiu Boeriu, *BMW to Collaborate With DeepSeek to Enhance In-Car AI*, BMWBLOG (Apr. 23, 2025), https://www.bmwblog.com/2025/04/23/bmw-deepseek-ai-china-integration [https://perma.cc/VS3D-VWWA].

[27] *See, e.g.*, Vincent Chow, *DeepSeek to Customise Namesake Chatbot for Italian Users Following Probe*, S. CHINA MORNING POST (Jan. 8, 2026, at 09:00 CT) https://www.scmp.com/tech/tech-trends/article/3339088/deepseek-customise-namesake-chatbot-italian-users-following-probe [https://perma.cc/MV4U-62KT].

[28] Tao Hong & Ming Hu, *Opportunities, Challenges, and Regulatory Responses to China's AI Computing Power Development under DeepSeek's Changing Landscape*, 2 INT'L J. DIGIT. L. & GOVERNANCE 83, 86–88 (2025); *see also* Anu Bradford, Eileen Li & Matthew C. Waxman, *How Domestic Institutions Shape the Global Tech War*, 16 HARV. NAT'L SEC. J. 75, 116 (2025) (indicating that DeepSeek has unveiled an AI model that operates with significantly less computing power than previously believed necessary); Mark A. Lemley & Matthew T. Wansley, *Coopting Disruption*, 105 B.U. L. REV. 457, 505 (2025) (noting that DeepSeek was reportedly trained at a comparatively low cost, coupled with its largely open access, presents a significant and potentially disruptive challenge to the established technology giants).

[29] Dan Steinfeld, *CHIPS, China and Choke Points: The Economic and National Security Consequences of U.S. Semiconductor Export Control Policy*, 57 N.Y.U. J. INT'L L. & POL. 456, 475 (2025).

Beyond Washington and Silicon Valley, DeepSeek-R1's ripple effect reverberated across Wall Street. DeepSeek's achievements signaled that

---

[30] Razii Abraham, *Democratizing AI's Frontiers: A Critical Review of DeepSeek AI's Open-Source Ecosystem*, OSF PREPRINTS (Feb. 6, 2025), https://osf.io/9r7e3_v1 [https://perma.cc/5J8B-4FMW]; *see also* Sheehan & Singer, *supra* note 6 (describing DeepSeek as an open-source generative AI model that rivals the most advanced offerings from U.S.-based OpenAI, achieving comparable capabilities at a fraction of the cost).

[31] *See, e.g.*, Cristina Criddle & Eleanor Olcott, *OpenAI Says It Has Evidence China's DeepSeek Used Its Model to Train Competitor*, FIN. TIMES (Jan. 29, 2025), https://www.ft.com/content/a0dfedd1-5255-4fa9-8ccc-1fe01de87ea6 [https://perma.cc/EJA4-LCVD]; David Goldman & Matt Egan, *A Shocking Chinese AI Advancement Called DeepSeek is Sending US Stocks Plunging*, CNN (Jan. 27, 2025, at 16:21 ET), https://www.cnn.com/2025/01/27/tech/deepseek-stocks-ai-china [https://perma.cc/KK4H-KU7W].

[32] Ng et al., *supra* note 8*; but see* Dan Steinfeld, Note *Chips, China and Choke Points: The Economic and National Security Consequences of U.S. Semiconductor Export Control Policy*, 57 N.Y.U. J. INT'L L. & POL. 456, 478 (2025) (noting that Dario Amodei, CEO of Anthropic, estimated that DeepSeek invested sums comparable to those of leading U.S. AI companies in developing a model with technical sophistication similar to that of U.S. models from seven to ten months prior).

[33] *See, e.g.*, Criddle & Olcott, *supra* note 31; Cade Metz, *What to Know About DeepSeek and How It Is Upending A.I.*, N.Y. TIMES (Jan. 27, 2025), https://www.nytimes.com/2025/01/27/technology/what-is-deepseek-china-ai.html [https://perma.cc/Q8JE-4QPK].

[34] Dennis Ronel, Note, *Regulating AI Power: How Law Can Save—Or Sink—U.S. Dominance*, 15 HARV. BUS. L. REV. 815, 817 (2025); Will Douglas Heaven, *How DeepSeek Ripped up the AI Playbook—and why Everyone's Going to Follow its Lead*, MIT TECH. REV. (Jan. 31, 2025), https://www.technologyreview.com/2025/01/31/1110740/how-deepseek-ripped-up-the-ai-playbook-and-why-everyones-going-to-follow-it [https://perma.cc/AP9W-AJ7K].

[35] *See, e.g.*, Wenjing Zhang et al., *Safety Evaluation of DeepSeek Models in Chinese Contexts*, CORN. U., May 8, 2025, at 2, arXiv:2502.11137v3 (indicating that in the 2024 American Invitational Mathematics Examination (AIME), DeepSeek-R1 achieved 79.8 percent accuracy, slightly outperforming OpenAI's o1, and on Codeforces, it outperformed 96.3 percent of human programmers.); *see also* DeepSeek-AI et al., *DeepSeek-R1: Incentivizing Reasoning Capability in LLMs via Reinforcement Learning*, CORN. U., Jan. 4 2026, at 13, arXiv:2501.12948v2 (reporting that based on LiveCodeBench results, DeepSeek-R1 outperforms Claude 3.5 Sonnet (1022), GPT-4o (0513), OpenAI o1-mini, and OpenAI o1 (1217)); Tianchen Gao et al., *A Comparison of DeepSeek and Other LLMs*, CORN. U., Dec. 29, 2025, at 1, arXiv:2502.03688v3 (finding that in terms of classification accuracy, DeepSeek generally outperforms Gemini, GPT, and Llama across most evaluations); Bakht Munir et al., *Evaluating AI in Legal Operations: A Comparative Analysis of Accuracy, Completeness, and Hallucinations in Chatgpt-4, Copilot, DeepSeek, Lexis+ Ai, and Llama 3*, 53 INT'L J. LEGAL INFO. 103, 108 (2025) (noting that DeepSeek generated more complete results than Copilot, Lexis+ AI, Llama 3, and ChatGPT-4).

America's preeminence in AI was under significant challenge, with China now emerging as a key global contender. NVIDIA, the leading provider of high-end chips, experienced a 12 percent drop in premarket trading[36]; the tech-heavy Nasdaq fell more than 3 percent in a broad sell-off that affected global chip makers and data centers.[37] In contrast, investor enthusiasm for Chinese technology companies surged,[38] propelling the Hang Seng Tech Index 25 percent above its January 2025 low,[39] while China's CSI 300 Index rose 4 percent.[40] Chinese tech giants including Alibaba, Xiaomi, Baidu, and BYD all enjoyed significant gains in their stock prices.[41]

DeepSeek points towards a potentially divergent trajectory for the future of AI development worldwide. Yet the profusion of commentary following its launch has often failed to pierce the veil of speculation, leaving many aspects of DeepSeek shrouded in mystery. In particular, the legal and policy implications of DeepSeek remain largely unexplored in current literature, while much of the media discourse is either speculative or inaccurate. This Article seeks to address this gap by providing rigorous legal and policy analysis relevant to the development of DeepSeek technologies.

Following this introduction, Part II examines the technical features of DeepSeek-R1, including its reinforcement learning mechanism, mixture-of-experts architecture, multi-head latent attention, distillation technology, and open-source nature, by discussing how: Reinforcement learning reduces reliance on supervised data by using preference signals and iterative self-refinement; The sparse mixture-of-experts routes tokens to a few specialized experts, expanding capacity without proportional compute; Multi-head latent attention structures attention over learned latent variables to improve retrieval and long-horizon reasoning; Distillation transfers capabilities from larger teachers to smaller students, preserving quality while lowering cost; and how DeepSeek's open-source practice releases code and, often, weights, enabling replication, auditing, and community-driven advances. Part III explores the major legal and policy issues surrounding DeepSeek, encompassing export control, information security, algorithmic bias, and intellectual property ("IP") infringement within a broader geopolitical debate by discussing how: Commentators have debated whether the model's very success challenges the efficacy of U.S. export restrictions aimed at curbing China's AI advancement; Regulators in various jurisdictions voice significant concerns over potential security vulnerabilities and embedded biases within the model; and how companies like OpenAI have publicly alleged that DeepSeek's achievements are predicated on the misappropriation of American IP. Part IV concludes.

---

[36] Goldman, *supra* note 12.

[37] Ng et al., *supra* note 8.

[38] Gregory C. Allen, *DeepSeek, Huawei, Export Controls, and the Future of the U.S.-China AI Race*, CTR. FOR STRATEGIC & INT'L STUD. (Mar. 7, 2025), https://www.csis.org/analysis/deepseek-huawei-export-controls-and-future-us-china-ai-race [https://perma.cc/BM8Q-9TB5].

[39] Arjun Neil Alim, *China's Tech Stocks Enter Bull Market After DeepSeek Breakthrough*, FIN. TIMES (Feb. 12, 2025), https://www.ft.com/content/af0636d6-59af-453c-9d75-6f6a3d9f11ce [https://perma.cc/4LLX-7K4C].

[40] *Id.*

[41] *Id.*

## II. DEEPSEEK TECHNOLOGIES

While the remarkable success of DeepSeek may have caught many by surprise, it is important to recognize that DeepSeek-R1 did not emerge out of thin air; rather, it is the culmination of years of dedicated technical development and innovation.[42] Prior to its groundbreaking appearance in the AI competition stage, the DeepSeek Company—established in 2023—had been actively refining and releasing successive versions of its models since November 2023,[43] steadily building a formidable foundation of expertise.[44] Many believe DeepSeek's distinctive technical features have enabled it to rival its American counterparts despite limited computing power and capital.[45]

Even before the success of DeepSeek-R1, the team had laid a strong foundation through a series of earlier models.[46] In January 2024, they introduced DeepSeek LLM,[47] an early exploration of scaling laws and data-model trade-offs.[48] In May 2024, they released the DeepSeek-V2 model, which adopted the innovative mixture-of-experts architecture and a multi-head latent attention mechanism[49]—advances that substantially improved the memory utilization and training efficiency of the model.[50] In December 2024, DeepSeek-V3 scaled sparse mixture-of-experts networks to 671 billion total parameters.[51] These developments were fully integrated into DeepSeek-R1, launched in

---

[42] *See* Daya Guo et al., *DeepSeek-R1: Incentivizing Reasoning Capability in LLMs via Reinforcement Learning*, 645 NATURE 633 (2025), https://www.nature.com/articles/s41586-025-09422-z [https://perma.cc/RH23-9QVH].

[43] Janakiram MSV, *All About DeepSeek — The Chinese AI Startup Challenging US Big Tech*, FORBES (Jan. 27, 2025, at 13:01 ET), https://www.forbes.com/sites/janakirammsv/2025/01/26/all-about-deepseekthe-chinese-ai-startup-challenging-the-us-big-tech [https://perma.cc/W82L-XBSP].

[44] Guo et al., *supra* note 42.

[45] *See, e.g.*, Ben Jiang, *DeepSeek Stays Mum on Next AI Model Release As Technical Papers Show Frontier Innovation*, S. CHINA MORNING POST (Jan. 14, 2026, at 10:00 CT), https://www.scmp.com/tech/tech-trends/article/3339769/deepseek-stays-mum-next-ai-model-release-technical-papers-show-frontier-innovation [https://perma.cc/6KPG-RRK5].

[46] Shayan Mohanty, *The DeepSeek Series: A Technical Overview*, MARTIN FOWLER (June 18, 2025), https://martinfowler.com/articles/deepseek-papers.html [https://perma.cc/W5MT-8KRT].

[47] *See generally* Xiao Bi et al., *DeepSeek LLM Scaling Open-Source Language Models with Longtermism*, CORN. U., Jan. 5, 2024, arXiv:2401.02954v1 (indicating that DeepSeek LLM was introduced in January of 2024).

[48] Mohanty, *supra* note 46.

[49] DeepSeek-AI et al., *DeepSeek-V2: A Strong, Economical, and Efficient Mixture-of-Experts Language Model,* CORN. U., Jun. 19, 2024, at 1 arXiv :2405.04434v5.

[50] Mohanty, *supra* note 46.

[51] DeepSeek-AI et al., *DeepSeek-V3 Technical Report*, CORN. U., Dec. 27, 2024, at 1, arXiv:2412.19437v1.

January 2025,[52] which also incorporates additional innovations such as large-scale reinforcement learning and knowledge distillation.[53]

DeepSeek's momentum did not stop with R1. In December 2025, the release of DeepSeek-V 3.2 further advanced the frontier of open large language models.[54] Its principal technical breakthroughs included DeepSeek Sparse Attention ("DSA"), a scalable reinforcement learning framework, and a large-scale agentic task synthesis pipeline.[55] Together, these cutting-edge technologies significantly enhanced the model's generalization capabilities and its instruction-following robustness in complex, interactive environments.[56]

This Part examines the key technical features that underpin DeepSeek's formidable AI performance. These defining characteristics include the use of reinforcement learning,[57] the adoption of a sophisticated mixture-of-experts architecture,[58] the implementation of a multi-head latent attention mechanism[59], the employment of knowledge distillation techniques,[60] and the embracing of an open-source software ("OSS") license.[61] These innovations have not only propelled DeepSeek to the forefront of AI excellence but have also sparked numerous legal controversies, such as the legality of knowledge distillation technology,[62] underscoring the profound impact and complexity of its technological underpinnings.

### A. *Reinforcement Learning*

Reinforcement learning technology refers to the process of training agents through a system of rewards and penalties, without specifying the exact pathways to complete a given task.[63] While DeepSeek is not the only model that employs reinforcement learning, it significantly refines the technique,

---

[52] Mohanty, *supra* note 46.

[53] *See generally* DeepSeek-AI et al., *supra* note 35 (indicating that both large-scale reinforcement learning and improved knowledge distillation were incorporated in DeepSeek-R1).

[54] *See generally* DeepSeek-AI et al., *DeepSeek-V3.2: Pushing the Frontier of Open Large Language Models*, CORN. U., Dec. 2, 2025, arXiv:2512.02556v1 (indicating that DeepSeek V 3.2 was released in December 2025).

[55] *Id.* at 1.

[56] *Id.*

[57] *See infra* Part II.A.

[58] *See infra* Part II.B.

[59] *See infra* Part II.C.

[60] *See infra* Part II.D.

[61] *See infra* Part II.E.

[62] *See, e.g.*, Brian Ferguson, Charles Liu & Yuchen Han, *Is AI Distillation by DeepSeek IP Theft?*, LAW360 (Mar. 7, 2025, at 14:55 ET), https://www.law360.com/articles/2306557 [https://perma.cc/7DWS-45Y4].

[63] Leslie Pack Kaelbling, Michael L. Littman & Andrew W. Moore, *Reinforcement Learning: A Survey*, 4 J. A.I. RSCH. 237, 237 (1996).

overcoming previous limitations and enhancing its overall effectiveness.[64] While the efficacy of reinforcement learning in reasoning tasks has long been established, the technology's reliance on vast quantities of supervised data has traditionally hampered the training efficiency of the base models.[65] However, DeepSeek-R1-0—the initial prototype of DeepSeek-R1—addresses this limitation by minimizing, and in some cases eliminating, the necessity for supervised data during reinforcement learning.[66] In other words, although DeepSeek-R1 is not the first model to use reinforcement learning, it significantly improves the training efficiency of the approach.

Central to this innovation is the adoption of group relative policy optimization, which enhances the reinforcement learning algorithm's efficiency. Rather than depending on a large and computationally intensive critic model, this approach estimates the baseline using group scores, thereby streamlining the optimization process.[67] Complementing this, DeepSeek-R1-0 employs a rule-based reward system that integrates two distinct types of incentives, accuracy rewards, which encourage correct solutions, and format rewards, which promote outputs adhering to specific structural guidelines.[68] Notably, to further conserve training resources and simplify the overall training pipeline, the model eschews the use of outcome-based or process-oriented neural reward mechanisms.[69]

Despite some early limitations—such as diminished readability and occasional language mixing—the foundational DeepSeek-R1 prototype paved the way for the more advanced DeepSeek-R1 model.[70] Through these substantial optimizations in the reinforcement learning mechanism, DeepSeek-R1 achieves markedly enhanced training efficiency, setting the new standard for the development of reasoning-capable AI systems.[71]

### B. *Mixture-of-Experts Architecture*

The mixture-of-experts ("MoE") architecture employed by the DeepSeek-R1 model plays a pivotal role in enabling its highly cost-effective performance. While the MoE architecture had already emerged as a prominent solution for mitigating the substantial computational demands of large models,[72] early implementations were hampered by issues such as knowledge hybridity and knowledge redundancy among experts.[73] DeepSeek-MoE addresses these challenges through the introduction of two core innovations: fine-grained expert segmentation and shared expert isolation.[74] By employing these strategies, DeepSeek-MoE not only surmounts the limitations of its predecessors but also achieves performance on par with prior models, such as DeepSeek 7B and LLaMA2 7B, all while requiring just 40 percent of the computational power.[75]

---

[64] Heaven, *supra* note 34 ("DeepSeek's use of reinforcement learning is the main innovation that the company describes in its R1 paper. But DeepSeek is not the only firm experimenting with this technique.").

[65] DeepSeek-AI et al., *supra* note 35, at 5.

[66] *Id.*

[67] *Id.*

[68] DeepSeek-AI et al., *supra* note 35, at 6.

[69] *Id.* at 6.

[70] *Id.* at 9.

[71] Sahin Ahmed, *DeepSeek-R1 explained: Pioneering the Next Era of Reasoning-Driven AI,* MEDIUM (Jan. 26, 2025), https://medium.com/@sahin.samia/deepseek-r1-explained-pioneering-the-next-era-of-reasoning-driven-ai-3eeb5ac4d4a0 [https://perma.cc/37U9-9HY9].

Since the launch of DeepSeek-R1, the team has made several advances to its MoE architecture. On December 31, 2025, it introduced Manifold-Constrained Hyper-Connections, addressing the training instability, limited scalability, and significant memory-access overhead seen in earlier hyper-connection designs.[76] On January 12, 2026, it released a paper detailing further MoE optimizations:[77] by adding a conditional memory module, Engram, the architecture resolves inefficient retrieval simulation.[78] Beyond more efficient retrieval, coupling MoE with Engram delivers gains in general reasoning, coding, and mathematics.[79] Mechanistic analysis indicates that Engram effectively deepens the network by freeing early layers from static reconstruction tasks, allowing the resulting attention capacity to focus on global context and more complex reasoning.[80] Treating infrastructure-aware efficiency as a first-class design principle, the team envisions conditional memory—exemplified by Engram—becoming an indispensable modeling primitive for next-generation sparse models.[81] Given its promise for boosting efficiency, some commentators suggest this approach offers a new training pathway to sidestep GPU constraints.[82]

---

[72] Damai Dai et al., *DeepSeekMoE: Towards Ultimate Expert Specialization in Mixture-of-Experts Language Models*, 2024 PROC. OF THE 62ND ANN. MEETING OF THE ASS'N FOR COMPUTATIONAL LINGUISTICS 1280, 1280 (2024); *see also* Saeed Masoudnia & Reza Ebrahimpour, *Mixture of Experts: A Literature Survey*, 42 A.I. REV. 275 (2014).

[73] Dai et al., *supra* note 72, at 1280.

[74] *Id.* at 1281–82.

[75] *Id.* at 1280–81.

[76] Zhenda Xie et al., *mHC: Manifold-Constrained Hyper-Connections*, CORN. U., Jan. 5, 2026, at 1, arXiv:2512.24880v2.

[77] *See generally* Xin Cheng et al., *Conditional Memory via Scalable Lookup: A New Axis of Sparsity for Large Language Models*, CORN. U., Jan. 12, 2026, arXiv:2601.07372v1.

[78] *Id.* at 1.

[79] *Id.* at 1, 20.

[80] *Id.* at 20–21.

[81] *Id.* at 21.

[82] Vincent Chow, *DeepSeek Founder's Latest Paper Proposes New AI Model Training to Bypass GPU Limits*, S. CHINA MORNING POST (Jan. 13, 2026 at 18:00 CT), https://www.scmp.com/tech/tech-trends/article/3339740/deepseek-founders-latest-paper-proposes-new-ai-model-training-bypass-gpu-limits [https://perma.cc/Z5KX-88JX]; Pan Xinghan, *DeepSeek Engram Explained: How Conditional Memory and O(1) Lookups Fix LLM Amnesia*, MEDIUM (Jan. 13, 2026), https://medium.com/@sampan090611/deepseek-engram-explained-how-conditional-memory-and-o-1-lookups-fix-llm-amnesia-50fce3937bb8 [https://perma.cc/Z36Y-SM8C].

### C. *Multi-Head Latent Attention Mechanism*

The attention mechanism serves to model the strength of relevance between pairs of representations.[83] Since its inception, Multi-Head Attention ("MHA")—which enables the simultaneous focus on information from multiple representation subspaces at different positions[84]—has achieved remarkable popularity, especially in transformers.[85] MHA has become the de facto architecture for natural language understanding tasks and neural machine translation.[86] While the standard MHA mechanism in transformer architecture relies on three distinct matrices for its computations, the multi-head latent attention ("MLA") approach introduces innovative compression and decompression matrices.[87] As a result, the DeepSeek-R1 model, equipped with MLA, is able to significantly reduce memory consumption without compromising the model's overall performance.[88]

### D. *Model Distillation*

Distillation stands out as one of DeepSeek's most notable technical features that garners significant attention and scrutiny. Commonly referred to as "knowledge distillation" in the field of computer science, this technique offers a promising solution to the challenges of computational complexity and substantial storage demands inherent in large-scale deep learning models.[89] Therefore, knowledge distillation has become a widely adopted strategy among researchers to create smaller, more efficient models that deliver performance nearly equivalent to much more computationally intensive counterparts.[90] It is important to acknowledge, however, that DeepSeek is not the originator of model distillation technology. Rather, DeepSeek has played a pivotal role in

---

[83] The attention mechanism can compare the degree of relevance between two words, such as "cow" and "grass." *See* Jian Li et al., *Multi-Head Attention with Disagreement Regularization*, 2018 PROC. CONF. ON EMPIRICAL METHODS NAT. LANGUAGE PROCESSING 2897, 2897–98 (2018).

[84] *Id.* at 2897.

[85] Jean-Baptiste Cordonnier, Andreas Loukas & Martin Jaggi, *Multi-Head Attention: Collaborate Instead of Concatenate*, CORN. U., May 20, 2021, at 1, arXiv:2006.16362v2.

[86] *Id.*

[87] Luolin Xiong et al., *DeepSeek: Paradigm Shifts and Technical Evolution in Large AI Models*, 12 IEEE/CAA J. AUTOMATICA SINICA 841, 847 (2025).

[88] *Id.*

[89] Jianping Gou et al., *Knowledge Distillation: A Survey*, 129 INT'L J. COMPUT. VISION 1789, 1789 (2021).

[90] *See, e.g.*, Peter Henderson & Mark A. Lemley, *The Mirage of Artificial Intelligence Terms of Service Restrictions*, 100 IND. L.J. 1327, 1367–68 (2025); Claudia Philipp, *From Prompt to Clone: Copyright Challenges in AI Model Distillation*, 17 UC L. SCI. & TECH. J. 49, 59 (2026); Criddle & Olcott, *supra* note 31.

demonstrating the transformative potential of this approach to a wider audience.[91]

The foundations of knowledge distillation can be traced back to 2015, when renowned computer scientists, Geoffrey Hinton, Oriol Vinyals, and Jeff Dean significantly improved the performance of a widely used commercial system "by distilling the knowledge in an ensemble of models into a single model."[92] This breakthrough underscored the capacity of model distillation to compress and transfer knowledge from complex, cumbersome models to more compact, efficient alternatives without substantial loss in accuracy.

At present, knowledge distillation generally falls into three categories: offline distillation, online distillation, and self-distillation.[93] To elucidate the underlying principles of this technique, researchers frequently draw an analogy to the education process—envisioning knowledge transfer as a skilled teacher imparting wisdom to a student.[94] In this framework, the "teacher model," often a complex and expansive network, generates a so-called "transfer set" by processing data with an elevated softmax temperature, yielding a soft target distribution. The "student model," typically more streamlined, is subsequently trained on this transfer set to assimilate the nuanced knowledge captured by the teacher.[95]

It is essential to recognize that a multitude of variables shape the efficacy and character of the "knowledge" transferred, such as the design of teacher-student architecture.[96] The success of distillation depends not just on the structures of the teacher and student models, but also on the nature of the knowledge being conveyed.[97] Forms of the "knowledge" distilled encompass response-based knowledge (the direct outputs or predictions), feature-based knowledge (internal representations), and relation-based knowledge (interconnections among data points or features).[98] This diversity has led to the development of numerous distillation algorithms tailored for varied scenarios:[99] for example, adversarial knowledge distillation seeks to align the student's grasp of the data distribution with that of the teacher; multi-teacher distillation integrates the knowledge, often in the form of logits [100] and feature representations sourced from multiple teachers; and teacher-guided architecture search allows the student model to emulate intermediate feature representations of the teacher at each search step. [101] The intricacies of distillation architecture and algorithm design create significant variability in the knowledge distilled and transferred from teacher model to student model—knowledge that is often referred to as "dark knowledge" due to its implicit and elusive nature.[102]

---

[91] Jasmine Wu & Deirdre Bosa, *How DeepSeek Used Distillation to Train Its Artificial Intelligence Model, and What it Means for Companies Such as OpenAI*, CNBC (Feb. 21, 2025, at 08:00 ET), https://www.cnbc.com/2025/02/21/deepseek-trained-ai-model-using-distillation-now-a-disruptive-force.html [https://perma.cc/RBM8-B4MG].

[92] Geoffrey Hinton, Oriol Vinyals & Jeff Dean, *Distilling the Knowledge in a Neural Network,* CORN. U., Mar. 9, 2015, at 1, arXiv:1503.02531v1.

[93] Gou et al., *supra* note 89, at 1795.

[94] Yuzhe Zhang et al., *LLM-Enhanced Multi-Teacher Knowledge Distillation for Modality-Incomplete Emotion Recognition in Daily Healthcare*, 29 IEEE J. BIOMEDICAL HEALTH INFORMATICS 6046, 6046 (2025).

[95] Hinton, Vinyals & Dean, *supra* note 92, at 1–3.

[96] Gou et al., *supra* note 89, at 1790, 1798.

[97] *Id.* at 1792.

[98] *Id.* at 1791.

[99] *Id.* at 1799–1804.

### E. *Open-Source Approach*

DeepSeek-R1 has been released under the Massachusetts Institute of Technology License ("MIT License"), [103] a permissive Open Source Software License ("OSS license") that allows licensees to use, modify, and distribute the code—including for commercial purposes—provided the original copyright and license notice are retained.[104] DeepSeek's distinction as one of the few truly open-source LLMs has garnered both acclaim and concern regarding various aspects of its underlying technology.[105]

On the positive side, commentators suggest that AI software released under OSS licenses—with source code made openly available—greatly fosters innovation, encourages community-driven development, and enhances accessibility, while simultaneously reducing costs.[106] DeepSeek's open-source strategy is widely regarded as exemplifying an "AI for all" philosophy, making technical advancement increasingly attainable for regions and institutes with limited resources.[107] As AI development accelerates and broadens, developers can focus their efforts on building specialized applications to solve real-world challenges.[108] Moreover, this openness enables users to customize models for tailored solutions[109] and offers greater transparency,[110] thereby helping to mitigate bias in DeepSeek's output. The transparency and accessibility inherent in open-source approaches also drive innovation far more rapidly than closed-door research.[111] For these reasons, some argue that OSS represents a more beneficial path for AI development.[112] There is even optimism among some observers that DeepSeek's success signals the potential for open-source to become the industry standard, or at the very least, to challenge the dominance of close-sourced models.[113]

---

[100] Essentially, logits are class-wise predictions. For instance, an image classification model calculates the likelihood of an input image belonging to each class it has been trained to identify and then selects the class with the highest probability. In machine learning terminology, these predictions for each class are referred to as logits. *See* Dave Bergmann, *What Is Knowledge Distillation?*, IBM, https://www.ibm.com/think/topics/knowledge-distillation [https://perma.cc/68P4-RCBH] (last visited Mar. 1, 2026).

[101] *Id.*

[102] *See, e.g.*, Utkarsh Ojha et al., *What Knowledge Gets Distilled in Knowledge Distillation?,* CORN. U., Nov. 6, 2023, at 2, arXiv:2205.16004v3; *see also* Geoffrey Hinton, Oriol Vinyals & Jeff Dean, *Dark Knowledge*, TTIC DISTINGUISHED LECTURE SERIES (2014), https://www.ttic.edu/dl/dark14.pdf [https://perma.cc/F77B-BDYM].

[103] Dylan Butts, *China's Open-Source Embrace Upends Conventional Wisdom Around Artificial Intelligence*, CNBC (Mar. 24, 2025, at 20:03 ET), https://www.cnbc.com/2025/03/24/china-open-source-deepseek-ai-spurs-innovation-and-adoption.html [https://perma.cc/WS7P-WP5Q].

[104] *See, e.g.*, Jyh-An Lee, *New Perspectives on Public Goods Production: Policy Implications of Open Source Software*, 9 VAND. J. ENT. & TECH. L. 45, 52–53 (2005).

[105] *Cf.* Lemley & Wansley, *supra note* 28 (describing the DeepSeek model as being a quasi-open-source one); Ronel, *supra* note 34 (noting that DeepSeek has adopted an "open-weight" model, allowing any researcher, startup, or competitor to modify, independently deploy, and build upon its innovations).

[106] *See, e.g.*, Charlotte Edmond, *What Is Open-Source AI and How Could DeepSeek Change the Industry?*, WORLD ECON. F. (Feb. 5, 2025), https://www.weforum.org/stories/2025/02/open-source-ai-innovation-deepseek [https://perma.cc/KQ3E-GYEP]; Alex Hern, *Meta Launches Open-Source AI App 'Competitive' with Closed Rivals*, GUARDIAN (July 23, 2024, at 14:47 ET), https://www.theguardian.com/technology/article/2024/jul/23/meta-launches-open-source-ai-app-

According to DeepSeek's API documentation, not only is the MIT License implemented, but the model's weight and output are also made available for community use.[114] Furthermore, API outputs can be employed for fine-tuning or distillation.[115] By contrast, while Meta's Llama-4 model—a major competitor to DeepSeek R1—professes a commitment to the OSS philosophy[116], its source

---

competitive-with-closed-rivals [https://perma.cc/67UN-NYXW]; Ankit Bisht et al., *Open Source in the Age of AI*, MCKINSEY & CO. (Feb. 11, 2025), https://www.mckinsey.com/capabilities/tech-and-ai/our-insights/tech-forward/open-source-in-the-age-of-ai [https://perma.cc/2QAS-WZ7S].

[107] Abraham, *supra* note 30 (noting that unlike proprietary labs that closely guard advanced models, DeepSeek releases most of its code and model weights (DeepSeek-AI, 2024c; DeepSeek-AI, 2025), fostering collaboration by enabling a flourishing ecosystem of experiments, replications, and derivative research—particularly for resource-constrained universities and independent researchers).

[108] *See, e.g.*, Janakiram MSV, *supra* note 43; Edmond, *supra* note 106.

[109] Xiaohan Xu et al., *A Survey on Knowledge Distillation of Large Language Models* (2024), CORN. U., Oct. 21, 2024, at 2, arXiv:2402.13116v1.

[110] *See, e.g.,* Md Naseef-Ur-Rahman Chowdhury et al., *DeepSeek vs. ChatGPT: A Comparative Analysis of Performance, Efficiency, & Ethical AI Considerations*, TECHRXIV (Feb. 11, 2025), https://www.techrxiv.org/users/690185/articles/1265410-deepseek-vs-chatgpt-a-comparative-analysis-of-performance-efficiency-and-ethical-ai-considerations?commit=bb94f295bdcf9a4f363a39f4a7a2472e950cbf39 [https://perma.cc/EFY3-JQYB]; *but see* Nizan Geslevich Packin & Karni Chagal-Feferkorn, *This Is Not a Game: The Addictive Allure of Digital Companions*, 48 SEATTLE U. L. REV. 693, 727 (2024) ("DeepSeek represents the latest iteration in a series of AI technologies that operate with minimal transparency and limited accountability").

[111] Wu & Bosa, *supra* note 91.

[112] Chow, *supra* note 6.

[113] Eric Schmidt & Dhaval Adjodah, *Will China's Open-Source AI End U.S. Supremacy in the Field?*, WASH. POST (Jan. 28, 2025), https://www.washingtonpost.com/opinions/2025/01/28/china-deekseek-ai-us-supremacy [https://perma.cc/7AB2-ZWL7].

[114] *DeepSeek API Docs*, DEEPSEEK (Jan. 20, 2025), https://api-docs.deepseek.com/news/news250120 [https://perma.cc/N8BP-J6BJ].

[115] *Id.*

[116] *The Llama 4 Herd: The Beginning of a New Era of Natively Multimodal AI Innovation*, META (Apr. 5, 2025), https://ai.meta.com/blog/llama-4-multimodal-intelligence [https://perma.cc/PNC2-KZ35].

code remains inaccessible to the public.[117] However, it is important to note that the adoption of an OSS license does not imply full transparency; only source code is disclosed, while the underlying training data remains unavailable.[118]

Conversely, the open-source nature of the DeepSeek model also raises concerns regarding information security vulnerability. Even before DeepSeek's release, questions surrounding the security of OSS-based AI models were being raised, as such models are perceived to be more susceptible to exploitation by malicious actors who may use them to spread misinformation[119] or comprise data privacy.[120] Unsurprisingly, the launch of DeepSeek-R1—granting users greater latitude to customize the model—has further intensified apprehensions about its susceptibility to cyber-attacks.[121] Critics warn that making powerful technology widely accessible and replicable may inadvertently facilitate its misuse, from the creation of bioweapons, large-scale phishing campaigns to the proliferation of low-quality AI-generated content.[122] These fears are substantiated by several technical evaluations: among others,[123] notably, Robust Intelligence and University of Pennsylvania conducted a safety assessment of DeepSeek-R1 alongside other frontier models—including Llama-3.1- 405B, GPT-4o, Gemini-1.5-pro, Claude-3.5-Sonnet, and o1-preview.[124] Using algorithmic jailbreaking techniques and applied automated attack methodologies, researchers were able to achieve a 100 percent Attack Success Rate in jailbreaking DeepSeek-R1, indicating the highest level of vulnerability among the models assessed.[125] It is noteworthy that Llama-3.1- 405B, the only other open-source model in the group, exhibited the second-highest ASR at 96 percent.[126]

---

[117] Shaun Sutner, *With Llama 4, Meta Ups Stakes in Open Model Race*, TECHTARGET (Apr. 10, 2025), https://www.techtarget.com/searchenterpriseai/news/366622436/With-Llama-4-Meta-ups-stakes-in-open-model-race [https://perma.cc/55PY-PQ4D].

[118] *See, e.g.*, Christian Vasquez, *DeepSeek Has Given Open-Source AI a Big Lift. But is it Safe for Companies to Use?*, FORTUNE (Feb. 4, 2025), https://fortune.com/2025/02/04/deepseek-open-source-ai-cybersecurity [https://perma.cc/S89A-5MCE].

[119] *See, e.g.*, Metz, *supra* note 33 (reporting that many experts warn that OSS could be easily "used to spread disinformation or cause other serious harm," and this issue has attracted some attention from the lawmakers).

[120] Vasquez, *supra* note 118.

[121] *CAISI Evaluation of DeepSeek AI Models Finds Shortcomings and Risks*, NAT'L INST. OF STANDARDS & TECH. (Sep. 30, 2025), https://www.nist.gov/news-events/news/2025/09/caisi-evaluation-deepseek-ai-models-finds-shortcomings-and-risks [https://perma.cc/UC5P-QUYW].

[122] Chow, *supra* note 6.

[123] *See, e.g.*, Sara Vera Marjanović et al., *DeepSeek-R1 Thoughtology: Let's Think about LLM Reasoning*, CORN. U., Jan. 15, 2026, at 32, arXiv:2504.07128v3 (finding that relative to other models, DeepSeek-R1 generates harmful responses to a substantial number of requests related to *Chemical & Biological Weapons*, *Cybercrime*, and *Misinformation*); *Zhang et al.*, *supra* note 35. (finding significant safety vulnerabilities associated with DeepSeek); *see also* Ctr. For Internet Sec. & Countering Hybrid Threats Team, *DeepSeek: A New Player in the Global AI Race*, CTR. FOR INTERNET SEC. (Mar. 24, 2025), https://www.cisecurity.org/insights/blog/deepseek-a-new-player-in-the-global-ai-race [https://perma.cc/8L7S-86KH] (noting that "researchers found several security vulnerabilities in DeepSeek that could expose sensitive data or lead to jailbreaking attempts, a technique intended to circumvent security measures").

[124] Paul Kassianik & Amin Karbasi, *Evaluating Security Risk in DeepSeek and Other Frontier Reasoning Models*, CISCO (Jan. 31, 2025), https://blogs.cisco.com/security/evaluating-security-risk-in-deepseek-and-other-frontier-reasoning-models [https://perma.cc/9L64-L3SL].

Nevertheless, the study revealed factors beyond DeepSeek's open-source nature that contribute to its security vulnerabilities. Researchers found that its susceptibility largely stems from cost-efficient training methods: cost savings in model development often result in compromised safety mechanisms.[127] Additional studies highlight that DeepSeek's Chain-of-Thought technology—which allows users to observe the model's step-by-step reasoning before receiving the final output[128]—may further expose the model to safety vulnerabilities.[129] Specifically, research has shown that fine-tuning attacks can exploit the intermediate reasoning process to generate highly detailed and harmful content.[130]

Despite the challenges and potential weakness inherent in OSS-based AI models, DeepSeek has successfully convinced the AI industry that OSS AI models presents a compelling alternative to LLM development and poses a serious challenge to its U.S. proprietary counterparts—most notably, OpenAI's ChatGPT and Google's Gemini.[131] Following DeepSeek's success, Chinese tech giants such as Alibaba, Baidu, and Tencent have also released their own OSS AI models.[132] In addition to DeepSeek's achievements, Alibaba's Qwen family of models has emerged as the world's largest open-source AI ecosystem, boasting over 100,000 derivative models—surpassing even the Llama community of Meta Platforms.[133]

Some commentators view open-source strategies as an effective means for Chinese LLM models to catch up, enabling broader collaboration and leveraging contributions from a global community of developers;[134] others argue that the OSS approach facilitates faster adoption and reduces development costs.[135] NVIDIA founder and CEO Jensen Huang has described China's open-source AI movement as a "catalyst for global progress," providing "every country and industry a chance to join the AI revolution."[136] Huang cited DeepSeek R1 to illustrate the accelerating momentum of open models in AI development—especially those released for broad use and adaptation—which, he said, have "reached the frontier."[137] The research team at Microsoft's AI for Good Lab also found that DeepSeek's open-source model has spurred broader AI adoption across the developing world, as its core components are freely available for anyone to access and modify.[138] As this vibrant ecosystem continues to mature, open-source AI is poised to transform the landscape of AI worldwide.

---

125 *Id.*; *See also* Matt Pearl, Julia Brock & Anoosh Kumar, *Delving into the Dangers of DeepSeek*, CTR. FOR STRATEGIC & INT'L STUD. (Feb. 24, 2025), https://www.csis.org/analysis/delving-dangers-deepseek [https://perma.cc/DMU3-QSXX] ("DeepSeek's design allows users to alter not only its functionalities but also its safety mechanisms, creating a far greater risk of exploitation. The absence of robust safeguards leaves the model exposed and makes it particularly vulnerable to jailbreaking, where attackers can bypass what little safety infrastructure exists to force the model to generate harmful content").

126 Kassianik & Karbasi, *supra* note 124.

127 *Id.*

128 Zhiyuan Xu, Joseph Gardiner & Sana Belguith, *The Dark Deep Side of DeepSeek: Fine-Tuning Attacks against the Safety Alignment of CoT-Enabled Models*, CORN. U., Feb. 3, 2025, at 2, arXiv:2502.01225v1.

129 Fengqing Jiang et al., *SafeChain: Safety of Language Models with Long Chain-of-Thought Reasoning Capabilities*, CORN. U., Feb. 17, 2025, at 1, arXiv:2502.12025v1.

130 Xu, Gardiner & Belguith, *supra* note 128, at 4.

131 Jiang, Feng & Chen, *supra* note 17.

132 Butts, *supra* note 103; Liam Mo & Brenda Goh, *China's Baidu to Make Latest Ernie AI Model Open-Source As Competition Heats Up*, REUTERS (Feb. 13, 2025, at 22:00 CT), https://www.reuters.com/technology/artificial-intelligence/baidu-make-ernie-ai-model-open-source-end-june-2025-02-14 [https://perma.cc/NGA5-9GFM].

133 Jiang, Feng & Chen, *supra* note 17.

## III. LEGAL AND POLICY IMPLICATIONS

Building on the preceding technical analysis of DeepSeek's technology, this Part examines the principal legal and policy issues associated with DeepSeek, namely, export control, information security, algorithmic bias, and intellectual property infringement.

### A. *Export Control*

Export control serves as both a policy instrument and a legislative tool, meticulously designed to safeguard national interests by regulating the export of specific goods and technologies to designated foreign nations.[139] These measures have become increasingly central amid the escalating technological and geopolitical rivalry between the United States and China, particularly

---

[134] *Id.*; Dashveenjit Kaur, *The China open-source AI revolution that's rattling Silicon Valley*, TECHWIRE ASIA (July 24, 2025), https://techwireasia.com/2025/07/china-open-source-ai-models-global-rankings/, [https://perma.cc/K7QE-YVWQ].

[135] Butts, *supra* note 103; *6 Reasons Why Open Source Software Lowers Development Costs*, THE LINUX FOUNDATION (Feb. 28, 2017), https://www.linuxfoundation.org/blog/blog/6-reasons-why-open-source-software-lowers-development-costs [https://perma.cc/6XD2-4386]; *see also* Siwen D. Cremean, *AI Export Controls: Safeguard or A Straitjacket?*, 40 BERKELEY TECH. L.J. 809, 856 (2025) (arguing that DeepSeek's OSS nature underpins its low cost and broad availability, which could confer a competitive edge internationally).

[136] Jiang, Feng & Chen, *supra* note 17.

[137] *See, e.g.*, *Nvidia CEO Spotlights DeepSeek as He Confirms Rubin AI Platform in Full Production*, XINHUA NEWS (Jan. 6, 2026), https://english.news.cn/20260107/62bdceaeda8640969708ca62a2bc110e/c.html [https://perma.cc/7PBU-355T].

[138] Chan Ho-Him & Matt O'Brien, *DeepSeek's AI Gains Traction in Developing Nations, Microsoft Report Says*, AP (Jan. 9, 2026), https://apnews.com/article/ai-deepseek-chatbot-china-microsoft-3ffc9b26f5798de8a7014fcd9bb343b0 [https://perma.cc/N86Q-C9HY].

[139] Cindy Whang, *Trade and Emerging Technologies: A Comparative Analysis of the United States and the European Union Dual-Use Export Control Regulations*, 31 SEC. & HUM. RTS. 11, 13 (2020).

within the AI sectors.[140]

Since October 2022, the U.S. Department of Commerce's Bureau of Industry and Security ("BIS") has intensified restrictions on the export of advanced computing chips, semiconductor chip manufacturing equipment, supercomputers, and integrated circuit ("IC") applications to China.[141] These measures were further escalated in October 2023,[142] April 2024,[143] December 2024,[144] and January 2025.[145] In each instance, the BIS justified these actions by citing U.S. national security, foreign policy objectives, and the imperative of maintaining technological supremacy—specifically highlighting China's strategic ambition to become a global AI leader by 2030.[146] In response, China retaliated by imposing its own export restrictions on critical minerals and rare earth elements ("REEs") fundamental to semiconductor production.[147]

---

140 Jyh-An Lee & Jingwen Liu, *Navigating Turbulence: The Challenges of Inclusive Innovation in the U.S.-China AI Race*, *in* INCLUSIVE INNOVATION IN THE AGE OF AI AND BIG DATA 322–29 (Daryl Lim & Peter Yu eds. 2026); Gregory C. Allen & Isaac Goldston, *Understanding U.S. Allies' Current Legal Authority to Implement AI and Semiconductor Export Controls*, CTR. FOR STRATEGIC & INT'L STUD. (Mar. 14, 2025), https://www.csis.org/analysis/understanding-us-allies-current-legal-authority-implement-ai-and-semiconductor-export [https://perma.cc/8LYA-XS65].

141 Implementation of Additional Export Controls: Certain Advanced Computing and Semiconductor Manufacturing Items; Supercomputer and Semiconductor End Use; Entity List Modification, 87 Fed. Reg. 62186 (Oct. 13, 2022) (to be codified at 15 C.F.R. pts. 734, 736, 740, 742, 744, 762, 772, and 774) [hereinafter Implementation of Additional Export Controls October 2022].

142 Implementation of Additional Export Controls: Certain Advanced Computing Items; Supercomputer and Semiconductor End Use; Updates and Corrections (AC/S IFR), 88 Fed. Reg. 73458 (Oct. 25, 2023) (to be codified at 15 C.F.R. pts. 732, 734, 736, 740, 742, 744, 746, 748, 758, 770, 772, and 774); Export Controls on Semiconductor Manufacturing Items (SME IFR), 88 Fed. Reg. 73424 (Oct. 25, 2023) (to be codified at 15 C.F.R. pts. 734, 736, 740, 742, 744, 772, and 774); Entity List Additions, 88 Fed. Reg. 71991 (Oct. 19, 2023) (to be codified at 15 C.F.R. pts. 744).

143 Implementation of Additional Export Controls: Certain Advanced Computing Items; Supercomputer and Semiconductor End Use; Updates and Corrections; and Export Controls on Semiconductor Manufacturing Items; Corrections and Clarifications, 89 Fed. Reg. 23878 (Apr. 4, 2024) (to be codified at 15 C.F.R. pts. 732, 734, 736, 740, 742, 744, 746, 748, 758, 770, 772, and 774).

144 Foreign-Produced Direct Product Rule Additions, and Refinements to Controls for Advanced Computing and Semiconductor Manufacturing Items, 89 Fed. Reg. 96790 (Dec. 5, 2024) (to be codified at 15 C.F.R. pts. 732, 734, 736, 740, 742, 744, 746, 758, 762, 772, and 774); Additions and Modifications to the Entity List; Removals from the Validated End-User (VEU) Program, 89 Fed. Reg. 96830 (Dec. 5, 2024) (to be codified at 15 C.F.R. pts. 744 and 748).

145 Framework for Artificial Intelligence Diffusion, 90 Fed. Reg. 4544 (Jan. 15, 2025) (to be codified at 15 CFR C.F.R. pts. 732, 734, 740, 742, 744, 748, 750, 762, 772, and 774).

146 Implementation of Additional Export Controls October 2022, *supra* note 141.

147 关于对镓、锗相关物项实施出口管制的公告 [Notice of the Gen. Admin. of Customs No. 23] (promulgated by the Gen. Admin. of Customs of the Ministry of Com., July 3, 2023, effective Aug. 1, 2023); 关于优化调整石墨物项临时出口管制措施的公告 [Notice of the Gen. Admin. of Customs No. 39] (promulgated by the Gen. Admin. of Customs of the Ministry of Com., Oct. 20, 2023, effective Dec. 1, 2023) [hereinafter *Notice on Graphite October 2023*]; 关于对锑等相关物项实施出口管制的公告 [Notice of the Gen. Admin. of Customs No. 33 [2024] (promulgated by the Gen. Admin. of Customs of the Ministry of Com., Aug. 15, 2024, effective Sep. 15, 2024) [hereinafter *Notice on Antimony August 2024*]; 关于加强相关两用物质对美国出口管制的公告[Notice of the Gen. Admin. of Customs No. 46] (promulgated by the Gen. Admin. of Customs of the Ministry of Com., Dec. 3, 2024, effective Dec. 3, 2024); 关于对部分中重稀土相关物项实施出口管制的决定 [Notice of the Gen. Admin. of Customs No. 18] (promulgated by the Gen. Admin. of Customs of the Ministry of Com., Apr. 4, 2025, effective Apr. 4, 2025).

Against this backdrop, the emergence of DeepSeek-R1 has sparked vigorous—and often divided—debates regarding the efficacy of U.S. export control policies. Some argue that restrictions on chip exports have failed to curb China's technological advancement;[148] instead, they have motivated China to explore alternative development pathways.[149] Describing DeepSeek's success as a "Sputnik moment" for AI reflects the view that China has already surmounted years of U.S. export restrictions and is now producing world-class AI models capable of competing on a global stage.[150] Consequently, there is a call for policymakers to reevaluate the continued validity of these controls.[151] Supporters of DeepSeek contend that the company's achievements demonstrate the potential to attain remarkable training efficiencies without dependence on "brute-force hardware acquisitions."[152] Such successes are viewed as evidence that external constraints could catalyze a paradigm shift in AI development.[153] This view is supported by DeepSeek's recently developed "conditional memory" technique, known as Engram, which addresses a key bottleneck in scaling AI models: the limited capacity of GPU high-bandwidth memory ("HBM").[154]

Most analysts, however, maintain that DeepSeek remains reliant on U.S.-origin chips. They interpret the company's achievements as exposing vulnerabilities within the current export regime.[155] Some suggest that U.S.

---

[148] Chow, *supra* note 6.

[149] Goldman, *supra* note 12; Sheehan, *supra* note 6; *see also* Steinfeld, *supra* note 29 (noting that some have claimed that Chinese AI development will not be significantly impeded by the use of Nvidia's slower chips, as algorithmic methods can overcome such hardware limitations—an assertion potentially exemplified by the release of the DeepSeek-R1 LLM in January 2025.).

[150] *See, e.g.*, Lee & Liu, *supra* note 140, at 328; Ronel, *supra* note 34, at 819.

[151] John Villasenor, *DeepSeek Shows the Limits of US Export Controls on AI Chips*, BROOKINGS (Jan. 29, 2025), https://www.brookings.edu/articles/deepseek-shows-the-limits-of-us-export-controls-on-ai-chips [https://perma.cc/777R-Z399] (concluding that "DeepSeek's models are a stark illustration of why U.S. export controls on advanced computing chips, instead of impeding China's AI progress, may actually be accelerating it.").

[152] Bradford, Li & Waxman, *supra* note 28, at 81 (stating that DeepSeek has recently challenged the prevailing notion that successful AI models must depend on high-performance chips); *see also* Cremean, *supra* note 135, at 855–56 (noting that DeepSeek's white papers demonstrate that it has developed a state-of-the-art AI system that is both cost- and compute-efficient); *DeepSeek's Achievements and the Implications for Policy*, NAT'L SEC. DATA & POL'Y INST. (Feb 21, 2025), https://nationalsecurity.virginia.edu/sites/default/files/files/2025-11/00006_20250221%29_nsdpi_deepseek%27s_achivements_and_the_implications_for_policy_3%29.pdf [https://perma.cc/ZVA3-CVRX].

[153] *See* NAT'L SEC. DATA & POL'Y INST., *supra* note 152.

[154] *See, e.g.*, *supra* text accompanying notes 76–82; Chow, *supra* note 84.

[155] Lennart Heim, *The Rise of DeepSeek: What the Headlines Miss*, HEIM: BLOG (Jan. 25, 2025), https://blog.heim.xyz/deepseek-what-the-headlines-miss [https://perma.cc/6SWP-LTFH]; Ashley Lin & Lennart Heim, *DeepSeek's Lesson: America Needs Smarter Export Controls*, RAND (Feb. 5, 2025), https://www.rand.org/pubs/commentary/2025/02/deepseeks-lesson-america-needs-smarter-export-controls.html [https://perma.cc/97J9-SC45]; *see also* Steve Hall, *Understanding DeepSeek's Impact on the AI Industry*, ISG, https://isg-one.com/articles/understanding-deepseek-s-impact-on-the-ai-industry [https://perma.cc/ZE9Z-BNQ5] (last visited Mar. 1, 2026) (arguing that "Additionally,

export controls operate with a significant time lag, making it premature to declare their failure.[156] Over recent years, as U.S. export restrictions have expanded, Chinese firms have adapted by stockpiling lower-end or older-generation hardware prior to restrictions tightening,[157] exemplified by reports that DeepSeek founder Liang Wenfeng began purchasing thousands of NVIDIA graphics processing units as early as 2021.[158] This perspective is substantiated by DeepSeek's recent postponement of its R2 model launch, a decision prompted by the company's inability to successfully train the model using Huawei's chips.[159] The challenges encountered by DeepSeek underscore the persistent gap between Chinese semiconductors and their American counterparts, particularly in executing complex and demanding tasks. This disparity vividly illustrates the obstacles China faces in its ambitious pursuit of technological self-sufficiency.[160]

The U.S. government acknowledged existing loopholes, as indirect supply channels through third countries like Singapore, Malaysia, and the United Arab Emirates facilitate Chinese access to U.S. chips.[161] This practice, colloquially termed "Singapore washing," undermines the effectiveness of export controls.[162] In response, the House Select Committee on the CCP has initiated an investigation into NVIDIA, demanding disclosure of all communications with

---

DeepSeek's reliance on older Nvidia chips highlights a key vulnerability, as U.S. export restrictions on advanced semiconductors to China present a growing challenge.").

[156] Heim, *supra* note 155; *see also* Cremean, *supra* note 135, at 855 (explaining that the effects of U.S.-imposed AI hardware controls may take time to become apparent).

[157] *See, e.g.*, *Chinese Imports of Chip Gear Hit Record $26 Billion This Year*, BLOOMBERG (Aug. 22, 2024), https://www.bloomberg.com/news/articles/2024-08-22/chinese-imports-of-chip-gear-hit-record-26-billion-this-year [https://perma.cc/A33Q-NBWC]; Iris Deng, *Tech War: China's Chip Imports Surge as Firms Stockpile Ahead of Fresh US Restrictions*, S. CHINA MORNING POST (Aug. 7, 2024), https://www.scmp.com/tech/tech-war/article/3273606/tech-war-chinas-chip-imports-surge-firms-stockpile-ahead-fresh-us-restrictions?module=perpetual_scroll_0&pgtype=article [https://perma.cc/R6SC-XW8C].

[158] Jason Ma, *Meet the Hedge Fund Manager Who Founded DeepSeek, the Chinese AI Startup That Began as a Hobby and is Now Laying Waste to U.S. Stocks*, FORTUNE (Jan. 27, 2025), https://fortune.com/2025/01/27/deepseek-founder-liang-wenfeng-hedge-fund-manager-high-flyer-quant-trading [https://perma.cc/CP5N-JD57]; *认识梁文峰：DeepSeek 爆火背后的对冲基金经理*[*Meet Liang Wenfeng: The Hedge Fund Manager Behind DeepSeek's Meteoric Rise*], 新浪科技 [SINA TECH.] (Jan. 28, 2025), https://finance.sina.com.cn/tech/roll/2025-01-28/doc-inehppua5164396.shtml [https://perma.cc/585N-QJTV].

[159] *See, e.g.*, Eleanor Olcott & Zijing Wu, *DeepSeek's Next AI Model Delayed by Attempt to Use Chinese Chips*, FIN. TIMES (Aug. 14, 2025), https://www.ft.com/content/eb984646-6320-4bfe-a78d-a1da2274b092 [https://perma.cc/5WMX-QDMU].

[160] *Id.*

[161] James Kynge, Jude Webber & Christine Murray, *"Singapore Washing": China's New Back Doors into Western Markets*, FIN. REV. (Sep. 9, 2024), https://www.afr.com/world/asia/singapore-washing-china-s-new-back-doors-into-western-markets-20240909-p5k8yj [https://perma.cc/3RFN-QYEW]; *see also* Cremean, *supra* note 135, at 856 (indicating that as long as alternative supply channels exist—whether through other countries, the black market, or pre–export-control stockpiles—they will continue to undermine U.S. export controls); Sheehan, *supra* note 6; (arguing that if Washington hopes to reclaim its leadership in frontier AI technologies, its first priority should be to close the existing gaps in the Commerce Department's export control policies).

[162] Kynge, Webber & Murray, *supra* note 161; *see also* Sheehan, *supra note* 6 (describing the smuggle of chips by Chinese companies to bypass the U.S. export controls).

DeepSeek and detailed information on transactions with customers in China (including Hong Kong and Macau) as well as ASEAN countries since 2020.[163]

It is important to note that export control policies are often influenced by the level of geopolitical tension and lobbying efforts. For example, on April 9, 2025, NVIDIA announced that the U.S. government had notified it of the inclusion of its H20 chips in export restrictions.[164] The H20 chips—similar to previously modified models such as the H800 chips and A800 chips—is a less advanced variant derived from NVIDIA's flagship H100 chips, adapted specifically for Chinese customers to circumvent restrictions.[165] Allegations suggest that the H20 chips significantly contributed to DeepSeek's early 2025 breakthroughs.[166]

The export ban on the H20 chips was finally lifted on July 16, 2025—roughly a month after high-level trade talks between the United States and China in London[167] —and the U.S. government subsequently erased restrictions on the sale of NVIDIA's H200 chip to China with a 25 percent surcharge in early 2026.[168] Many observers interpret the U.S. move to lift export control on chips as aimed at increasing dependency on U.S. technology, with NVIDIA poised to be the primary benefactor of the relaxation.[169] Even so, Chinese firms have not placed orders after Beijing encouraged them to pivot to domestic suppliers such as Huawei.[170]

In sum, export controls continue to be a dynamic element of the broader strategic contest, shaped by geopolitical tensions, industry lobbying,[171] and evolving technological landscapes.[172]

---

[163] *Nvidia Letter (Mr. Jensen Huang, CEO)*, H.R. SELECT COMM. ON THE CCP (Apr. 16, 2025), https://chinaselectcommittee.house.gov/media/letters/nvidia-letter-mr-jensen-huang-ceo [https://perma.cc/J4Q9-JKS8].

[164] NVIDIA Corp., *Current Report Pursuant to Section 13 or 15(d) of the Security Exchange Act of 1934*, SEC. & EXCH. COMM'N (Apr. 9, 2025), https://www.sec.gov/ix?doc=/Archives/edgar/data/0001045810/000104581025000082/nvda-20250409.htm. [https://perma.cc/VNG2-ZWM5].

[165] Liam Mo & Brenda Goh, *Exclusive: Nvidia modifies H20 chip for China to overcome US export controls, sources say*, REUTERS (May 9, 2025, at 04:21 CT), https://www.reuters.com/world/china/nvidia-modifies-h20-chip-china-overcome-us-export-controls-sources-say-2025-05-09 [https://perma.cc/V4VE-MRV9].

[166] *See* Section III.1.B.

[167] 黄仁勋：将开始向中国市场销售 H20 芯片 [*Jensen Huang: Will Commence Sales of H20 Chips to the Chinese Market*], CCTV NEWS (July 14, 2025), https://content-static.cctvnews.cctv.com/snow-book/index.html?item_id=10051782561450980212&t=1752546250770&toc_style_id=feeds_default &share_to=wechat&track_id=cd320e5b-a206-4b19-8914-c52414acb09e [https://perma.cc/MQX4-795Q].

[168] *See, e.g.*, Megha Shrivastava, *Nvidia's H200 Chips Re-enter China – But Beijing Isn't Giving Up on Huawei*, DIPLOMAT (Feb. 3, 2026), https://thediplomat.com/2026/02/nvidias-h200-chips-re-enter-china-but-beijing-isnt-giving-up-on-huawei/

[169] *See, e.g.*, Dylan Butts, *Nvidia's Set to Regain Some China Access. But It Still Faces Eroding AI Chip Market Share*, CNBC (Aug. 4, 2025, at 08:53 ET), https://www.cnbc.com/2025/08/04/nvidia-h20-china-market-share-recovery.html [https://perma.cc/6S56-R8EE].

[170] *See, e.g.*, Miranda Jeyaretnam, *How A.I. Was the Elephant in the Room at the Trump-Xi Summit*, TIME (May 15, 2026), https://time.com/article/2026/05/15/trump-xi-us-china-summit-ai-semiconductor-chips/.

### B. *Information Security*

Information security has long been a prevalent concern surrounding LLMs. For example, in 2023, employees at Samsung inadvertently leaked sensitive data and code to ChatGPT while using the tool for work-related purposes.[173] Consequently, OpenAI, the creator of ChatGPT, gained access to this information, as its privacy policy permits the collection of user-provided data, including prompts and uploaded files.[174] In response to growing security worries, Samsung banned its staff from using external AI applications and began developing its own in-house artificial intelligence system.[175]

Nevertheless, the effectiveness of such bans in mitigating information security risks remains uncertain, particularly given the rising prevalence of "Shadow AI," or unauthorized use of AI in the workplace.[176] Data leaks can occur not only to model developers but also to third parties. For example, a technical vulnerability at the South Korean company AI-Nomis led to the exposure of thousands of users' prompts and AI-generated images, making this information accessible to anyone on the Internet.[177] Similarly, Wiz Research, a New York–based cybersecurity firm, found that DeepSeek had inadvertently left a ClickHouse database unprotected online, a misconfiguration that granted unrestricted access to sensitive information, including chat histories, API keys, and backend operational details.[178]

Beyond the general apprehensions related to LLMs, however, DeepSeek's Chinese origin—coupled with the current geopolitical climate—presents a unique dimension of information security risk. This distinct layer of concern quickly manifested following the release of DeepSeek's R1 model, which was met with prohibitions in several countries on the grounds of safeguarding information security concern. For example, on January 30, 2025, Italy blocked access to the DeepSeek model and forbid the company from processing personal data of Italian citizens.[179] On February 3, 2025, Taiwan barred government departments from using DeepSeek AI over information security reasons.[180] The following day, Australia banned DeepSeek on government devices, citing security risk concerns; Australian officials emphasized that the decision was based on the "unacceptable risk" posed to national security, rather than the model's Chinese origin.[181] Later in February, South Korea removed DeepSeek from app stores to address privacy-related security concerns.[182]

---

[171] Bradford, Li & Waxman, *supra* note 28, at 103 ("DeepSeek's recent AI breakthrough may invite greater scrutiny of corporate influence, in particular due to bipartisan accusations that Biden-era export control loopholes promoted by industry lobbying enabled it.").

[172] *See, e.g.*, Jingwen Liu & Jyh-An Lee, *Strategic Stalemates: The Paradox of Export Controls in the U.S.-China AI Race*, 35 J. TRANSNAT'L L. & POL'Y 152, 201 (2026).

[173] Mark Gurman, *Samsung Bans Staff's AI Use After Spotting ChatGPT Data Leak*, BLOOMBERG (May 2, 2023), https://www.bloomberg.com/news/articles/2023-05-02/samsung-bans-chatgpt-and-other-generative-ai-use-by-staff-after-leak [https://perma.cc/8QYS-KFAF].

[174] *Privacy Policy*, OPENAI, https://openai.com/policies/row-privacy-policy [https://perma.cc/B75W-38QN] (last visited Feb. 28, 2026, at 18:56 CT).

[175] Gurman, *supra* note 170.

[176] *See generally*, Deepak Puthal et al., *Shadow AI: Cyber Security Implications, Opportunities and Challenges in the Unseen Frontier*, 6 SN COMPUT. SCI. 1 (2025).

[177] Matt Burgess, *An AI Image Generator's Exposed Database Reveals What People Really Used It For*, WIRED (Mar. 31, 2025, at 06:00 CT), https://www.wired.com/story/genomis-ai-image-database-exposed [https://perma.cc/98D3-Q3XF].

[178] Munir et al., *supra* note 35, at 108; Thomas Claburn, *Guess Who Left a Database Wide Open, Exposing Chat Logs, API Keys, and More? Yup, DeepSeek*, THE REGISTER (Jan. 30, 2025 00:31 UTC), https://www.theregister.com/2025/01/30/deepseek_database_left_open [https://perma.cc/MX62-9DEA]; Kevin Poireault, *DeepSeek Exposed Database Leaks Sensitive Data*, INFOSECURITY MAGAZINE (Jan. 30, 2025), https://www.infosecurity-magazine.com/news/deepseek-database-leaks-sensitive [https://perma.cc/ZV46-VYAJ]

In the United States, both federal and state governments—as well as the military—have implemented a series of bans on the use of DeepSeek. For instance, as early as January 28, 2025, the U.S. Navy issued a warning to its personnel, advising that DeepSeek should not be used "in any capacity" due to "potential security and ethical concerns."[183] On January 31, 2025, Texas Governor Greg Abbott announced a prohibition on Chinese AI and social media applications, including DeepSeek, on government-issued devices,[184] making Texas the first state to enact such a ban.[185] On February 10, 2025, New York State followed suit, prohibiting DeepSeek on government devices.[186] Governor Kathy Hochul expressed serious concerns regarding "apparent censorship" and "foreign government surveillance."[187] At the federal level, the U.S. Department of Commerce banned DeepSeek,[188] notifying staff in March 2025 that the model was prohibited on all government-furnished equipment.[189] Alongside legislative and executive measures in a dozen states that restrict certain government officials from installing DeepSeek on state devices, twenty-one state attorneys general sent a letter urging Congress to enact the proposed federal "No DeepSeek on Government Devices Act."[190]

---

[179] Ng et al., *supra* note 8.

[180] *Taiwan Bans Government Departments Using DeepSeek AI*, REUTERS (Feb. 3, 2025, at 05:25 CT), https://www.reuters.com/technology/taiwan-bans-government-departments-using-deepseek-ai-2025-02-03 [https://perma.cc/4ASW-DLXV].

[181] Tom Gerken, *Australia Bans DeepSeek on Government Devices Over Security Risk*, BBC (Feb. 4, 2025), https://www.bbc.com/news/articles/c8d95v0nr1yo [https://perma.cc/MY5S-P7XK].

[182] João da Silva & Jean Mackenzie, *S Korea Removes DeepSeek from App Stores Over Privacy Concerns*, BBC (Feb. 17, 2025), https://www.bbc.com/news/articles/clyzym0vn8go [https://perma.cc/3ELW-7GBU].

[183] Hayden Field, *U.S. Navy Bans Use of DeepSeek Due to 'Security and Ethical Concerns'*, CNBC NEWS (Jan. 28, 2025, at 17:16 ET), https://www.cnbc.com/2025/01/28/us-navy-restricts-use-of-deepseek-ai-imperative-to-avoid-using.html [https://perma.cc/ZD7N-JWXH]

[184] *Governor Abbott Announces Ban on Chinese AI, Social Media Apps*, OFF. TEX. GOVERNOR (Jan. 31, 2025), https://gov.texas.gov/news/post/governor-abbott-announces-ban-on-chinese-ai-social-media-apps [https://perma.cc/J5TA-NBHX].

[185] *See* Caitlin McCormack, *Texas Becomes First State to Ban DeepSeek, Rednote on Government Devices After Fury Over China-Backed Apps*, N.Y. POST (Feb. 2, 2025), https://nypost.com/2025/02/02/us-news/texas-becomes-first-state-to-ban-deepseek-rednote-for-government-devices [https://perma.cc/DQ98-WHJP].

[186] *Governor Hochul Issues Statewide Ban on DeepSeek Artificial Intelligence for Government Devices and Networks*, N.Y. STATE (Feb. 10, 2025), https://www.governor.ny.gov/news/governor-hochul-issues-statewide-ban-deepseek-artificial-intelligence-government-devices-and [https://perma.cc/34QJ-4YX4].

Many of these concerns stem from the perception that the Chinese government exerts significant control over information held by Chinese companies. Commentators have noted that, in the absence of due process and an independent judiciary, the Chinese government could arbitrarily access data maintained by private enterprises.[191] Despite the existence of vague legal provisions governing government access to private data, critics maintain that, because DeepSeek stores its user data within China's borders,[192] Chinese

---

[187] Angela Yang, *New York State Bans DeepSeek from Government Devices*, NBC NEWS (Feb. 11, 2025, at 14:22 CT), https://www.nbcnews.com/tech/new-york-state-bans-deepseek-government-devices-rcna191510 [https://perma.cc/RY9U-7384].

[188] Karen Freifeld, *US Commerce Department Bureaus Ban China's DeepSeek on Government Devices, Sources Say*, REUTERS (Mar. 18, 2025, at 18:23 CT), https://www.reuters.com/technology/artificial-intelligence/us-commerce-department-bureaus-ban-chinas-deepseek-government-devices-sources-2025-03-17 [https://perma.cc/4J7P-B2CN].

[189] *Id.*

[190] NBC Montana, *Attorney General Knudsen leads 21-state push to ban China-based AI on government devices*, NBC MONTANA(Mar. 8, 2025), https://nbcmontana.com/news/local/attorney-general-knudsen-leads-21-state-push-to-ban-china-based-ai-on-government-devices; Jamie Joseph, *It's time to ban Chinese AI app DeepSeek from 'government devices,' state AGs urge Congress*, FOX NEWS(Mar.6, 2025), https://www.foxnews.com/politics/its-time-ban-chinese-ai-app-deepseek-from-government-devices-state-ags-urge-congress.

[191] *See, e.g.*, Jyh-An Lee, *Hacking into China's Cybersecurity Law*, 53 WAKE FOREST L. REV. 57, 101–03 (2018); *see also* Karman Lucero, *Artificial Intelligence Regulation and China's Future*, 33 COLUM. J. ASIAN L. 94, 151 (2019) (warning that the Chinese government can access personal and enterprise data through its social credit system, often without due process).

[192] China has established a comprehensive legal framework for the protection of personal data protection. In 2021, the Data Security Law of the People's Republic of China—widely regarded as the cornerstone of the nation's data security regime—came into effect. Together with the Cyber Security Law and the Personal Information Protection Law, these three laws delineate the regulatory landscape for data governance in China. Moreover, all three laws specify the circumstances under which Chinese law enforcement agencies may request data sharing, detailing the perquisite conditions, procedures and the liabilities incurred when due process is not observed. For example, Article 45 of the Cyber Security Law states that, "[t]he departments assuming cybersecurity supervision and administration functions in accordance with the law and their staff members shall strictly keep confidential the personal information, privacy and trade secrets to which they have access in performing their functions, and shall not divulge, sell or illegally provide the information to any other person." *See* 中华人民共和国网络安全法 [Cyber Security Law] (promulgated by Standing Comm. Nat'l People's Cong., Nov. 7, 2016, revised Oct. 28, 2025, effective Jan. 1, 2026), art. 45, https://www.pkulaw.com/en_law/fe9682732480b562bdfb.html [https://perma.cc/DZ6T-MEET]. Article 68 of Personal Information Protection Law stipulates that, "[w]here any state organ fails to fulfill the personal information protection obligations as provided in this Law, the superior organ or the authority performing personal information protection functions shall order it to take corrective action, and take disciplinary action against the directly liable person in charge and other directly liable persons in accordance with the law. Where any

intelligence agencies possess the legal authority to compel access to this information without safeguards of due process.[193] This situation raises serious concerns about privacy and the potential for state surveillance, given the opacity of the mechanisms through which such demands might be enforced.[194] An even graver worry is that the DeepSeek model could be co-opted as a tool for government espionage, enabling surveillance of DeepSeek's users.[195] Such distrust has only been intensified amid escalating tensions between China and the United States, as well as the growing institutionalization of Communist Party organizations within private companies.[196]

These suspicions are by no means unique to DeepSeek. They extend to all Chinese information platforms seeking to expand into Western markets, such as

---

functionary of the authority performing personal information protection functions neglects duty, abuses power, or makes falsification for private gain, if it is not criminally punishable, he or she shall be subject to disciplinary action in accordance with the law." *See* 中华人民共和国个人信息保护法 [Personal Information Protection Law] (promulgated by Standing Comm. Nat'l People's Cong., Aug. 20, 2021, effective Nov. 1, 2021), art. 68, http://www.npc.gov.cn/npc/c2/c30834/202108/t20210820_313088.html [https://perma.cc/U34N-4G3A]. Article 35 of Data Security Law stipulates that, "[a]s needed for maintaining national security or investigating crimes, a public security authority or national security authority shall legally pull data in accordance with relevant provisions issued by the state and by strictly following approval procedures, and the relevant organizations and individuals shall provide cooperation."

*See* 中华人民共和国数据安全法 [Data Security Law] (promulgated by Standing Comm. Nat'l People's Cong., June 10, 2021, effective Sep. 1, 2021), art. 35, http://www.npc.gov.cn/npc/c2/c30834/202106/t20210610_311888.html [https://perma.cc/YZZ6-Y6Y3].

[193] Pearl, Brock & Kumar, *supra* note 125 ("The platform's Terms of Service state that DeepSeek is "governed by the laws of the People's Republic of China in the mainland." DeepSeek's Privacy Policy states that user data is stored in the PRC and governed by PRC law. Given that PRC law mandates cooperation with PRC intelligence agencies, these policies provide the PRC with great flexibility to access DeepSeek user data without the legal process that would be required in a rule-of-law country.").

[194] *See* Michael Martina & Stephen Nellis, *Exclusive: DeepSeek Aids China's Military and Evaded Export Controls, US Official Says*, REUTERS (June 23, 2025, at 09:49 CT), https://www.reuters.com/world/china/deepseek-aids-chinas-military-evaded-export-controls-us-official-says-2025-06-23 [https://perma.cc/6WWD-AKES]; CIS Countering Hybrid Threats Team, *DeepSeek: A New Player in the Global AI Race*, THE CTR. FOR INTERNET SEC. (Mar. 24, 2025), https://www.cisecurity.org/insights/blog/deepseek-a-new-player-in-the-global-ai-race [https://perma.cc/QZZ2-EMC8].

[195] *See, e.g.*, Pearl, Brock & Kumar, *supra* note 125. (contending that " DeepSeek is the latest tool in the PRC's cyber espionage toolkit to obtain more comprehensive intelligence and support the country's strategic and geopolitical objectives.").

[196] *See, e.g.*, Christopher Chen & Lauren Yu-Hsin Lin, *Foreign Political Interference in the Governance of Listed Companies: A Market and Behavioral Analysis*, 31 DUKE J. COMP. & INT'L L. 301, 307 (2021) (contending that China has mandated that listed companies incorporate party organization provisions into their corporate constitutions, heightening the risk that the Chinese Communist Party may exert influence over privately held and even foreign enterprises); Lauren Yu-Hsin Lin & Curtis J. Milhaupt, *Party Building or Noisy Signaling? The Contours of Political Conformity in Chinese Corporate Governance*, 50 J. LEGAL STUD. 187, 215 (2021) ("The corporate governance of Chinese public companies appears to grow considerably more complex as political considerations are formally introduced into corporate decision-making and personnel").

TikTok and Huawei.[197] For Chinese technology companies, no amount of robust data protection or practices can fully dispel these concerns,[198] which are deeply rooted in geopolitical realities and the structure of China's governing system.[199] China's distinctive governance model often enables the party-state to strategically mobilize commercial entities in pursuit of Beijing's foreign policy objectives.[200] This seamless integration of political authority and private enterprise empowers the government to exert influence beyond its borders through ostensibly independent companies. As a result, concerns regarding the presence of Chinese firms in international markets stem from their deep entanglement with state power,[201] with many believing that these companies are ultimately obliged to support government censorship and surveillance efforts.[202] Therefore, it is unsurprising that similar concerns have surfaced in relation to DeepSeek.

Nonetheless, while these apprehensions may appear credible on the surface, to date there has arguably been no concrete evidence demonstrating that the Chinese government has accessed foreign data through these corporations.[203] Even among those who believe that China has collected personal data on Americans, there is agreement that no evidence exist showing that China has derived any tangible benefit from such activities.[204] The controversy and widespread distrust of China's data protection regime extend beyond legal statutes and their practical enforcement; they have emerged from a complex

---

[197] Chow, *supra* note 6; *see also* Packin & Chagal-Feferkorn, *supra* note110, at 727 ("[b]uilding on the lessons learned from TikTok's controversial trajectory, the emergence of China's DeepSeek AI model in early 2025 has amplified global potential concerns about data usage and management, privacy, manipulation, and much more"); Robert D. Williams, *Beyond Huawei and TikTok: Untangling U.S. Concerns over Chinese Tech Companies and Digital Security* (Brooking, Working Paper, 2020) (on file with the Penn Project on the Future of U.S.-China Relations) (indicating that concerns regarding Chinese technology companies operating in the United States typically encompass not only the technologies themselves but also the risks inherent to the nature of China's governing system).

[198] Chow, *supra* note 6.

[199] Williams, *supra* note 194.

[200] Aziz Z. Huq, *Book Review: The Geopolitics of Digital Regulation*, 92 U. CHI. L. REV. 833, 894 (2025).

[201] ANU BRADFORD, DIGITAL EMPIRES: THE GLOBAL BATTLE TO REGULATE TECHNOLOGY 293 (2023).

[202] *Id.*; Jyh-An Lee & Ching-Yi Liu, *Searching for Internet Freedom in China: A Case Study on Google's China Experience*, 31 CARDOZO ARTS & ENT. L.J. 405, 426–28 (2013).

[203] *See, e.g.*, Max Zahn, *No Evidence of TikTok National Security Threat but Reason for Concern, Experts Say*, ABC NEWS, (Mar. 28, 2023, at 11:07 CT), https://abcnews.go.com/Technology/evidence-tiktok-national-security-threat-reason-concern-experts/story? [https://perma.cc/68ES-464A]; Rachel Sandler, *CIA Finds 'No Evidence' Chinese Government Has Accessed TikTok Data, Report Says*, FORBES (Aug. 7, 2020, at 20:11 ET), https://www.forbes.com/sites/rachelsandler/2020/08/07/cia-finds-no-evidence-chinese-government-has-accessed-tiktok-data-report-says [https://perma.cc/33NU-PH8E].

[204] *TikTok and National Security*, CTR. FOR STRATEGIC & INT'L STUD. (Mar. 13, 2024), https://www.csis.org/analysis/tiktok-and-national-security [https://perma.cc/LQV6-RGVX] ("China does collect personal data on Americans and have been doing so for at least a decade, but there is no evidence that it has found a way to benefit from this.")

interplay of economic, political, cultural factors.[205]

### C. *Algorithmic Bias*

While AI algorithms have emerged as remarkably efficient and effective tools for identification, assessment, and content creation, many remain susceptible to bias introduced during data collection or processing.[206] Such biases can lead to prejudiced decisions and the generation of content shaped by demographic characteristics, such as race, ethnicity, gender, or religion.[207] These biases may, therefore, give rise to legal challenges concerning discrimination across various areas of law.[208] Some scholars contend that these biases represent one of the most urgent and significant challenges facing contemporary society.[209] The origins of algorithmic bias can often be traced to the quality and representativeness of data used in training,[210] the particular machine learning techniques employed,[211] and the fundamental design of the algorithms themselves.[212]

Algorithmic bias is a widespread concern across numerous AI models,[213] and DeepSeek is no exception.[214] In particular, the model's Chinese origin has raised specific doubts regarding bias in its output: notably linguistic[215] and political biases.[216] Linguistic bias arises when an LLM is trained predominantly on data from a single language, causing it to favor that language and its associated culture or worldview.[217] This can result in content that fails to accurately reflect the diversity of human experiences and languages.[218] A recent quantitative assessment found that DeepSeek exhibits a 21.7 percent disparity in safety performance between Chinese and English contexts; notably, prompts submitted in English are more likely to elicit potentially harmful content.[219]

---

[205] *See, e.g.*, Jyh-An Lee, *Algorithmic Bias and the New Chicago School*, 14 L. INNO & TECH. 95–112 (2022) (using privacy and free speech as examples to explain that human rights in China diverge fundamentally from those in the Western world: they are never conceived as affording individuals rights that supersede those of the Chinese state); *see also* Mark Jia, *Authoritarian Privacy*, 91 U. CHI. L. REV. 733, 737 (2024) (noting that "[p]rivacy violators... [in China] include opportunistic criminals, avaricious firms, and sloppy local governments, but never the central party-state itself); Williams, *supra* note 194.

[206] Lee, *supra* note 202, at 95.

[207] *Id.*

[208] *See, e.g.*, Sandra G. Mayson, *Bias in, Bias out*, 128 YALE L.J. 2218, 2218 (2019); Cass R. Sunstein, *Governing by Algorithm? No Noise and (Potentially) less Bias*, 71 DUKE L.J. 1175 (2022); Emily Black et al., Solon Baroc, *Less Discriminatory Algorithms*, 113 GEO. L.J. 53 (October 2024); Spencer M. Mainka, *Algorithm-Based Recruiting Technology in the Workplace*, 5 TEX. A&M J. PROP. L. 801 (2019); Goodwin Liu, *Implicit Bias, Structural Bias, and Implications for Law and Policy*, 25 U. PA. J. CONST. L. 1280 (2023); W. Nicholson II Price, *Medical AI and Contextual Bias*, 33 HARV. J.L. & TECH. 65 (2019); Elizabeth Pendo & Jennifer D. Oliva, *Disability Discrimination by Clinical Algorithm*, 103 N.C. L. REV. 187 (2024); Wendy K. Tam Cho & Bruce E. Cain, *Deploying Trustworthy AI in the Courtroom: Lessons from Examining Algorithm Bias in Redistricting AI*, 2023 U. CHI. LEGAL F. 87 (2023).

[209] *See, e.g.*, Lee, *supra* note 202, at 95–96; Harry Surden, *Artificial Intelligence and Law: An Overview*, 35 GA. ST. U. L. REV. 1305, 1335 (2019).

[210] Solon Barocas & Andrew D. Selbst, *Big Data's Disparate Impact*, 104 CAL. L. REV. 671 ( 2016).

[211] Konstantinos Mavrogiorgos et al., *Bias in Machine Learning: A Literature Review*, 14 APPLIED SCI. 8860 (2024).

[212] Judith Simon, Pak-Hang Wong & Gernot Rieder, *Algorithmic Bias and the Value Sensitive Design approach*, 9 INTERNET POL'Y REV. 1 (2020).

[213] *See, e.g.* Fabio Motoki, Valdemar Pinho Neto & Victor Rodrigues, *More Human than Human: Measuring ChatGPT Political Bias*, 198 PUB. CHOICE 3, 3 (2024) (finding that ChatGPT is biased towards "the Democrats in the US, Lula in Brazil, and the Labor Party in the UK"); Partha Pratim Ray, *ChatGPT: A Comprehensive Review on Background, Applications, Key Challenges, Bias, Ethics, Limitations and Future Scope*, 3 INTERNET OF THINGS AND CYBER-PHYSICAL SYS. 121, 147–48 (2023) (identifying various biases in ChatGPT's output).

Even more concerning is the presence of ideological or political bias within DeepSeek's models, which sometimes overlap with the above-mentioned linguistic bias.[220] Such bias occurs when an LLM reflects only the dominant perspectives present in its training data.[221] For instance, when users seek information on controversial topics, the model may provide responses aligned solely with prevailing political, social, or economic ideologies, thereby reinforcing existing biases and creating an unbalanced representation of different opinions.[222] Commentators have criticized DeepSeek's models for being subject to censorship intended to prevent criticism of the Chinese Communist Party, posing a significant obstacle to global adoption.[223] For example, the DeepSeek chatbot reflects Chinese censorship, declining to address topics such as the 1989 Tiananmen Square event.[224] In countries where freedom of expression is highly valued, this form of censorship may diminish DeepSeek's appeal and acceptance.[225] By contrast, DeepSeek thrives where American technology is curtailed, posting its highest estimated market shares in China (89 percent), Belarus (56 percent), Cuba (49 percent), Russia (43 percent), and Iran (25 percent).[226]

---

214 Zehang Deng et al., *Exploring DeepSeek: A Survey on Advances, Applications, Challenges and Future Directions*, 12 IEEE/CAA J. AUTOMATICA SINICA 872, 886 (2025).

215 Waleed Kadous, *DeepSeek is Amazing. And it Has a Pro-Chinese Bias*, MEDIUM (Jan. 28, 2025), https://waleedk.medium.com/deepseek-is-amazing-and-it-has-a-pro-chinese-bias-78e2fd8e40bb [https://perma.cc/YP42-UDWL].

216 Ian Reynolds, Benjamin Jensen & Yasir Atalan, *Hawkish AI? Uncovering DeepSeek's Foreign Policy Biases*, CTR. FOR STRATEGIC & INT'L STUD. (Apr. 16, 2025).

217 Ray, *supra* note 210, at 147.

218 *Id.*

219 Zonghao Ying et al., *Towards Understanding the Safety Boundaries of DeepSeek Models: Evaluation and Findings* (Mar. 19, 2025) (unpublished manuscript) (on file with the University of Iowa Journal of Transnational Law and Contemporary Problems).

220 Di Zhou & Yinxian Zhang, *Political Biases and Inconsistencies in Bilingual GPT Models—The Cases of the U.S. and China*, 14 SCI. REPORTS 25048 (2024).

221 Ray, *supra* note 210, at 147.

222 Moshe Glickman & Tali Sharot, *How Human–AI Feedback Loops Alter Human Perceptual, Emotional and Social Judgements*, 9 NATURE HUM. BEHAVIOR 345, 345 (2025).

223 *See, e.g.*, Sheehan & Singer, *supra* note 6; (indicating that Some observers fear that DeepSeek's success suggests China may be better positioned to propagate AI models that restrict free speech and censor inconvenient truths—such as those relating to Tiananmen Square and Taiwan—that challenge its leaders' political objectives).

224 *Which Countries Are Adopting AI the Fastest?*, ECONOMIST (Jan. 12, 2026, at 17:53 CT), https://www.economist.com/graphic-detail/2026/01/12/which-countries-are-adopting-ai-the-fastest [https://perma.cc/M66J-8G8N].

The Center for Strategic and International Studies ("CSIS") conducted a comparative analysis of foreign policy tendencies in several prominent models—including Llama 3.1 8B Instruct, Llama 3.1 70B Instruct, GPT-4o, Gemini 1.5 Pro-002, Mixtral 8x22B, Claude 3.5, and Qwen2 72B, and DeepSeek V3—using the CSIS Futures Lab Critical Foreign Policy Decision Benchmark.[227] Their findings indicate that DeepSeek displays a distinct inclination towards hawkish and escalatory recommendations.[228] Other researchers noted that "[DeepSeek] R1 exhibits a unique form of censorship behavior aligned with specific political sensitivities,"[229] marking "the first instance where a locally deployed version of a model exhibits local censorship behavior."[230] These allegations of political bias have emerged as one of the major criticisms of DeepSeek in the international market.[231]

Nevertheless, we argue that the linguistic and political biases embedded in DeepSeek are unlikely to threaten the company's competitive position in the short term. First, similar political biases exist among DeepSeek's Western counterparts. For example, studies have shown that ChatGPT often produces left-leaning political statements.[232] Another study demonstrates that ChatGPT favors certain parties or actors including the Democrats in the United States, Lula in Brazil, and the Labor Party in the UK, rather than presenting a comprehensive array of viewpoints.[233] Therefore, political bias is not unique to DeepSeek; however, geopolitical tensions may amplify concerns about such biases, hindering DeepSeek's global adoption.

---

[225] Janakiram MSV, *supra* note 43.

[226] *Which Countries Are Adopting AI the Fastest?*, *supra* note 221.

[227] Reynolds, Jensen & Atalan, *supra* note 213.

[228] *Id.*

[229] Ali Naseh et al., *R1dacted: Investigating Local Censorship in DeepSeek's R1 Language Model*, CORN. U., May 19, 2025, at 13, arXiv:2505.1265v1.

[230] *Id.* at 3.

[231] *See, e.g.*, Roni Berkowitz & Yael Ram, *What You (Don't) See is What You Get: DeepSeek and the Myth of Technical Neutrality*, LONDON SCH. ECON. AND POL. SCI. BLOG (Aug. 25, 2025), https://blogs.lse.ac.uk/cff/2025/08/25/what-you-dont-see-is-what-you-get-deepseek-and-the-myth-of-technical-neutrality [https://perma.cc/EM6L-LAKM]; Angela Yang, *On DeepSeek, You Can Watch AI Navigate Censorship in Real Time*, NBC NEWS (Jan. 30, 2025, at 05:00 ET), https://www.nbcnews.com/tech/innovation/deepseek-censorship-china-rcna189594 [https://perma.cc/S53N-L2YX]; Mary Whitfill Roeloffs, *Does DeepSeek Censor Its Answers? We Asked 5 Questions on Sensitive China Topics*, FORBES (Jan. 27, 2025, at 15:28 ET), https://www.forbes.com/sites/maryroeloffs/2025/01/27/does-deepseek-censor-its-answers-we-asked-5-questions-on-sensitive-china-topics [https://perma.cc/Y5Q5-SFS5].

[232] David Rozado, *The Political Biases of ChatGPT*, 12 SOC. SCI. 148, 151 (2023); Sara Harrison, *Study Finds Perceived Political Bias in Popular AI Models*, STAN. REP. (May 21, 2025), https://news.stanford.edu/stories/2025/05/ai-models-llms-chatgpt-claude-gemini-partisan-bias-research-study [https://perma.cc/TB55-QTF6]; Jeremy Baum & John Villasenor, *The Politics of AI: ChatGPT and Political Bias,* BROOKINGS INST. (May 8, 2023), https://www.brookings.edu/articles/the-politics-of-ai-chatgpt-and-political-bias [https://perma.cc/93JV-JGKU].

Second, DeepSeek's models have demonstrated superior or comparable performance in specific domains, such as Chinese natural language processing,[234] coding,[235] and mathematics.[236] A study published on Nature Medicine concludes that DeepSeek models perform as well as or even surpass GPT-4o and Gemini-2.0 in an experiment involving 125 patient cases that covered a broad spectrum of common and rare diseases.[237] Additionally, DeepSeek's open-source approach—which enables onsite deployment—ensures data privacy and compliance with healthcare regulations.[238] In specific areas like complex ophthalmology reasoning, DeepSeek R1 has outperformed other state-of-the-art models.[239] Coupled with its advantage in affordable API pricing,[240] it is reasonable to anticipate that DeepSeek will attract users who prioritize domain-specific capabilities, local deployability, and cost efficiency. Thus, for users whose expectations are not politically oriented, DeepSeek's criticized political biases may have little impact on their acceptance of the model.

In light of these considerations, this Article argues that while algorithmic bias underlies DeepSeek's models, it is unlikely to drive them out of the market given the prevalence of similar problems among competitors and DeepSeek's strength in specialized domains. However, should DeepSeek aim to further expand its global reach, it will inevitably need to address these issues, demonstrating both the comprehensiveness of its training data and the neutrality in its outputs.

### D. *IP and Contract*

The rapid success of DeepSeek has provoked criticisms and allegations from the U.S. government and American enterprises, who claim that the company's achievements are founded upon infringement of IP belonging to its American counterpart.[241] During his nomination hearing before the Senate Commerce, Science, and Transportation Committee, Secretary of Commerce Howard Lutnick testified that DeepSeek had managed to develop a powerful AI model "dirt cheap" by leveraging stolen U.S. technology.[242] Similarly, President Donald Trump's AI and crypto adviser, David Sacks, indicated that IP theft by DeepSeek was a distinct possibility.[243] The Select Committee on the Chinese Communist Party ("CCP"), in its report, concludes that it is highly probable DeepSeek employed "unlawful model distillation techniques" to copy the leading U.S. models.[244] OpenAI has likewise asserted that "DeepSeek may have inappropriately distilled our models . . . We take aggressive, proactive countermeasures to protect our technology and will continue working closely with the US government to protect the most capable models being built here."[245] Yet, while the report declared DeepSeek's knowledge distillation techniques as unlawful, it provided few details about the alleged infringement.[246] In February 2026, OpenAI submitted a memorandum to the Congress asserting that they had observed accounts associated with DeepSeek employees developing methods to circumvent OpenAI's access restrictions, including by using obfuscated third-party routers and other techniques intended to mask their origin.[247] In February 2026, Anthropic—another U.S. AI company—also alleged that three Chinese AI firms—DeepSeek, Moonshot AI, and MiniMax—had engaged in industrial-scale IP theft, using roughly 24,000 fake accounts to conduct about 16 million interactions with its Claude chatbot in order to illicitly extract its capabilities.[248] Responding to Lutnick's claims, Liu Pengyu, spokesperson for Beijing's embassy in Washington, countered that the United States had failed to offer compelling evidence to support its accusations of IP theft.[249] Although DeepSeek has faced allegations of distilling knowledge from proprietary models such as ChatGPT,[250] the company maintains that its distillation efforts are limited exclusively to open-source models, including Qwen and Llama.[251]

---

[233] Motoki, Neto & Rodrigues, *supra* note 210, at 3 (finding that ChatGPT is biased towards "the Democrats in the US, Lula in Brazil, and the Labor Party in the UK").

[234] Swiss German University, *A Comparison of Leading AI Models: DeepSeek AI, ChatGPT, Gemini, and Perplexity AI*, SWISS GERMAN UNIV. BLOG (Feb. 7, 2025), https://sgu.ac.id/a-comparison-of-leading-ai-models-deepseek-ai-chatgpt-gemini-and-perplexity-ai [https://perma.cc/2ERV-2VFG].

[235] *See, e.g.*, Mihika Sharma, *DeepSeek to Launch New AI Model Focused on Coding in February, The Information Reports*, REUTERS (Jan. 9, 2026, at 08:33 CT) https://www.reuters.com/technology/deepseek-launch-new-ai-model-focused-coding-february-information-reports-2026-01-09 [https://perma.cc/5QU3-7852].

[236] Lee & Liu, *supra* note 140, at 302; DeepSeek-AI et al., *supra* note 35, at 13.

[237] Sarah Sandmann et al., *Benchmark Evaluation of DeepSeek Large Language Models in Clinical Decision-Making*, 31 NATURE MED. 2546, 2546 (2025).

[238] *Id.*

[239] *See generally,* Pusheng Xu et al., *DeepSeek-R1 Outperforms Gemini 2.0 Pro, OpenAI O1, and O3-Mini in Bilingual Complex Ophthalmology Reasoning*, 5 ADVANCES OPHTHALMOLOGY PRAC. & RSCH. 189 (2025).

[240] *See, e.g.*, Janakiram MSV, *supra* note 43.

---

[241] *See, e.g.*, Henderson & Lemley, *supra* note 90, at 1333 (noting that the similarities between the outputs of the DeepSeek R1 model and those of OpenAI's O1 model have led to accusations of "intellectual property theft").

[242] *See, e.g.*, *DeepSeek Leveraged US Chips, "Stolen" Technology, Trump's Commerce Secretary Pick Says*, REUTERS (Jan. 29, 2025, at 14:09 CT), https://www.reuters.com/world/us/trumps-commerce-secretary-pick-says-biden-era-chips-investments-need-review-2025-01-29 [https://perma.cc/VLS4-PHSF]; *see also* Robert Delaney, *Trump's Commerce Secretary Pick Vows to Keep China From Using US Tech to 'Compete With Us'*, S. CHINA MORNING POST (Jan. 30, 2025, at 06:05 CT), https://www.scmp.com/news/china/article/3296743/trumps-commerce-secretary-pick-vows-keep-china-using-us-tech-compete-us [https://perma.cc/NA7Q-5WW9] (quoting Lutnick's statement that "[Chinese AI companies] stole things. They broke in. They've taken our IP and it's got to end."); Ryan Lovelace, *'Outrageous': Commerce Pick Lutnick Lashes Out at China's Suspected AI Theft for DeepSeek*, WASH. TIMES (Jan. 29, 2025), https://www.washingtontimes.com/news/2025/jan/29/commerce-secretary-nominee-howard-lutnick-lashes-c [https://perma.cc/CF7V-JH4H] (reporting that Lutnick described DeepSeek as an IP theft whose conduct was outrageous).

[243] *See, e.g.*, Criddle & Olcott, *supra* note 31.

[244] The Select Comm. on the CCP, *DeepSeek Unmasked: Exposing the CCP's Latest Tool for Spying, Stealing, and Subverting U.S. Export Control Restrictions* (Apr. 16, 2025), https://selectcommitteeontheccp.house.gov/media/reports/deepseek-unmasked-exposing-ccps-latest-tool-spying-stealing-and-subverting-us-export [https://perma.cc/XQL8-GY74].

[245] Kevin Collier & Jasmine Cui, *OpenAI Says DeepSeek May Have "Inappropriately" Used Its Data*, NBC NEWS (Jan. 30, 2025, at 09:09 CT), https://www.nbcnews.com/tech/tech-news/openai-says-deepseek-may-inapproriately-used-data-rcna189872 [https://perma.cc/5TBS-GR93].

[246] The Select Comm. on the CCP, *supra* note 241.

[247] Deepa Seetharaman & Fabiola Arámburo, *OpenAI says China's DeepSeek trained its AI by distilling US models, memo shows*, REUTERS (Feb. 13, 2026, at 16:16 CT), https://www.reuters.com/world/china/openai-accuses-deepseek-distilling-us-models-gain-advantage-bloomberg-news-2026-02-12 [https://perma.cc/WL2L-Z9R3].

The question of whether knowledge distillation constitutes infringement remains largely underexplored within IP law. The central challenge is that, in the knowledge distillation process, the student model does not replicate IP-protected source code or arguably other proprietary information from the teacher model; indeed, most crucial protected information is never accessible to the student model. This section will focus on trade secrets and copyright—the two domains of IP law most often raised in relation to knowledge distillation—and will also examine whether the owner of the teacher model can claim that the developer of the student model has violated the former's terms of use.

### 1. Trade Secret

Some argue that trade secret law could offer OpenAI a potential avenue to hold DeepSeek liable for its use of knowledge distillation.[252] The U.S. Defend Trade Secrets Act ("DTSA") defines a trade secret as any form or type of information that its owner has taken reasonable steps to protect, and that derives independent economic value from not being generally known or readily ascertainable through proper means.[253] To prevail in claims for trade secret misappropriation, a plaintiff must first identify, with reasonable particularity, the specific information alleged to constitute a trade secret, and then demonstrate that this information is genuinely secret, possesses independent economic value, and has been subject to reasonable efforts to preserve its confidentiality.[254]

---

[248] *See* Agence France-Presse, *US AI Giant Accuses Chinese Rivals of Mass Data Theft*, GUARDIAN (Feb. 23, 2026, at 23:15 GMT), https://www.theguardian.com/technology/2026/feb/23/us-ai-anthropic-china.

[249] Delaney, *supra* note 242.

[250] John Werner, *Did DeepSeek Copy Off of OpenAI? And What is Distillation?*, FORBES (Jan. 30, 2025, at 16:10 ET), https://www.forbes.com/sites/johnwerner/2025/01/30/did-deepseek-copy-off-of-openai-and-what-is-distillation [https://perma.cc/PM2Q-XG44].

[251] DeepSeek-AI et al., *supra* note 35, at 14; *see also* Camilla A. Hrdy, *Keeping ChatGPT a Trade Secret While Selling It Too*, 40 BERKELEY TECH. L.J. 75, 113 (noting that DeepSeek has refuted these claims, maintaining that their model was independently developed).

[252] *See, e.g.*, Camilla A. Hrdy, *Trade Secrecy Meets Generative AI*, 100 CHI.-KENT L. REV. 317, 318, 349–50 (2025).

[253] *See* Defend Trade Secrets Act of 2016 § 2(b)(1), Pub. L. No. 114-153, § 1839(3), 130 Stat. 376, 380 (codified as amended at 18 U.S.C. § 90).

[254] *See, e.g.*, Mallet and Co. Inc. v. Lacayo, 16 F.4th 364 (3d Cir. 2021) (holding that a preliminary injunction based on trade secret misappropriation must clearly and specifically identify the trade secrets that are claimed to have been misappropriated).

If the plaintiff fails to distinguish the trade secret from other information, the claim will be rendered too vague and overly broad to be properly evaluated under the criteria of secrecy, economic value, and reasonable efforts.[255] Without precise identification, it is impossible to compare the alleged trade secret against information in the public domain.[256] For example, in *IDX Sys. Corp. v. Epic Sys. Corp.*, the plaintiff submitted a 43-page description of the methods, processes, and interrelationships comprising its software package.[257] The court found this broad description failed to separate the alleged trade secrets from other information, preventing a determination of which aspects were generally known within the industry and which were not.[258] Given that clear identification is a prerequisite under the statutory definition, the court declined to address the remaining elements of secrecy, value, and reasonable efforts.[259] Similarly, in *Perlan Therapeutics, Inc. v. Superior Court*, the court held that where a plaintiff cannot specify the disputed trade secrets, any of its allegations of misappropriation are baseless.[260]

This requirement of reasonable particularity or specificity poses a significant challenge for the proprietor of a teacher model seeking to establish trade secret misappropriation against the developer of the student model in the context of knowledge distillation. Identifying exactly what the student model has appropriated from the teacher model is an extraordinary complex undertaking. As previously described, knowledge distillation process is a singular process within AI:[261] the student model observes and analyzes the inference patterns manifested in the teacher model's output and then seeks to emulate these patterns. Through such learning, the student model internalizes the teacher's manner of reasoning and generates output that resembles that of the teacher when presented with the same or similar input.

The inherent "black box" nature of AI algorithms further complicates this inquiry, rendering it nearly impossible to trace or reconstruct the original inference path used by the teacher model.[262] In practice, the designer of the student model independently selects and adjusts the parameters in the distilling algorithm—for example, various intermediate features.[263] Utilizing these parameters, the student model charts its own unique inference path, mapping inputs to outputs and ultimately producing results akin to those of the teacher model.[264] Variations in design choices inevitably yield differences in both training efficiency and reasoning processes.[265] Thus, multiple student models may extract distinct forms of what is sometimes called "dark knowledge" from a single teacher model.[266]

---

[255] *See generally* IDX Sys. Corp. v. Epic Sys. Corp., 285 F.3d 581 (7th Cir. 2002) (indicating that the plaintiff-appellant failed to distinguish what it claimed to be a trade secret from other general software information rendering the claim too vague to pursue).

[256] Charles Tait Graves & Brian D. Range, *Identification of Trade Secret Claims in Litigation: Solutions for Ubiquitous Dispute*, 5 NW. J. TECH. & INTELL. PROP. 68, 68–69 (2006); *see also* Jyh-An Lee & Lauren Yu-Hsin Lin, *Legal Institutions of Human Capital: Property, Contract, and Organisation*, 13 J. KNOWLEDGE ECON. 3326 (2022) (noting that case law indicates that a trade secret must be distinguished from other information that the employee is entitled to use freely).

[257] *IDX Sys. Corp.*, 285 F.3d at 583–84.

[258] *Id.*

[259] *IDX Sys. Corp.*, 285 F.3d 581.

[260] Perlan Therapeutics, Inc. v. Super. Ct. of San Diego Cnty., 101 Cal. Rptr. 3d 211, 218–19 (Cal. Ct. App. 2009).

[261] *See supra* Part II. D.

In essence, the designers of the student model are not merely copying the output of the teacher model; rather, they are carefully observing and analyzing the teacher's output, striving to imitate its voice as captured in its outputs. This process is highly variable—much like the proverbial notion that "there are a thousand Hamlets in a thousand people's eyes," the knowledge extracted by a thousand student model designers from the same teacher model may take a thousand different forms. Without a clear identification of what constitutes "distilled knowledge," it becomes fundamentally difficult—if not impossible—for creators of distilled models such as OpenAI to ascertain any specific trade secret allegedly misappropriated.

Furthermore, while it is widely acknowledged that AI models are composed of several elements that may be protected as trade secrets—including the model's algorithm,[267] model weights,[268] source code,[269] training data,[270] optimization techniques,[271] and system architecture[272]—the process of studying a teacher model's output during knowledge distillation typically does not enable the student model to appropriate any of these elements.[273] Because the teacher model's output is intentionally made available to users, its disclosure precludes itself from trade-secret protection.[274] Moreover, the output alone provides insufficient access to the underlying components of the teacher model, further complicating any claim of trade secret misappropriation.[275]

---

[262] *See, e.g.*, Cynthia Rudin, *Stop Explaining Black Box Machine Learning Models for High Stakes Decisions and Use Interpretable Models Instead*, 1 NATURE MACH. INTEL. 206 (2019); Vikas Hassija et al., *Interpreting Black-Box Models: A Review on Explainable Artificial Intelligence*, 16 COGNITIVE COMPUTATION 45, 45 (2023).

[263] Ojha et al., *supra* note 102, at 1 (noting that "[t]he seminal work by demonstrated its effectiveness by making the student imitate the teacher's class probability outputs for an image. This ushered in an era of knowledge distillation algorithms, including those that try to mimic intermediate features of the teacher, or preserve the relationship between samples as modeled by the teacher, among others.")

[264] *Id.*

[265] Lucas Beyer et al., *Knowledge Distillation: A Good Teacher Is Patient and Consistent*, 2022 IEEE/CVF CONF. ON COMPU. VISION & PATTERN RECOGNITION (CVPR) 10915, 10915 (2022), https://ieeexplore.ieee.org/document/9879513 [https://perma.cc/A2LV-6GNT].

[266] Hinton, Vinyals & Dean, *supra* note 102.

Although we do share the view of some scholars who characterize knowledge distillation as a form of reverse engineering,[276] we nonetheless believe that the way trade secret law addresses reverse engineering provides illuminating guidance for issues related to knowledge distillation.

---

267 *See, e.g.*, LivePerson Inc. v. 24/7 Customer Inc., 83 F. Supp. 3d 501, 514 (S.D.N.Y. 2015); Hrdy, *supra* note 251, at 96–99; Gregory Gerard Greer, *Artificial Intelligence and Trade Secret Law*, 21 UIC REV. INTELL. PROP. L. 252, 263 (2022).

268 *See, e.g.*, Ridoan Karim, *AI Distillation and the Law: Why Learning From Claude or GPT May Not Be Copyright Infringement,* MONASH LENS (March 16, 2026), https://lens.monash.edu/ai-distillation-and-the-law-why-learning-from-claude-or-gpt-may-not-be-copyright-infringement/.

269 Hrdy, *supra* note 251, at 100–03. *See also* Hunton Andrews Kurth LLP, *A Practical Guide to Protecting AI Models with Trade Secrets, Fed. Bar Council Quarterly,* HUNTON (Aug. 1, 2025), https://www.hunton.com/insights/publications/a-practical-guide-to-protecting-ai-models-with-trade-secrets [https://perma.cc/7NNM-CCXG].

270 Hrdy, *supra* note 251, at 103–06. *See also* Suppakit Waiwitlikhit et al., *Trustless Audits without Revealing Data or Models*, CORN. U., Apr. 6, 2024, at 1, arXiv.2404.04500v1 (noting that "[a]s ML has become more capable and is being deployed in sensitive settings, models and data are increasingly kept proprietary. This is because data and model weights are increasingly treated as trade secrets.")

271 Karim, *supra* note 268.

272 Hrdy, *supra* note 251, at 106–08; Sam Mitchell, Nicole Bashor & Heather Ehlers, *Protecting AI Innovation: Why Trade Secrets are Outpacing Patents in IP Portfolios*, 72 RISK MGMT. MAG. 8, 9 (2025); Indictment at 8, United States v. Linwei Ding, CASE NO. 3:24-cr-00141 VC (N. D. Cal. Mar. 5, 2024) (https://www.justice.gov/archives/opa/media/1341356/dl?inline) [https://perma.cc/WT97-4G3L]. According to the indictment, the trade secrets stolen by the former Google employee include, chip architecture and software design specifications for TPU version 4, chip architecture and software design specifications for TPU version 6, hardware, software, system management, and performance specifications for GPU chips deployed in Google's supercomputing data Centers, Software design specifications for Google CMS that managed machine learning workloads on TPU and GPU chip in Google's supercomputing data centers; Press Release, U.S. Department of Justice, Chinese National Residing in California Arrested for Theft of Artificial Intelligence-Related Trade Secrets from Google (Mar. 6, 2024) (https://www.justice.gov/archives/opa/pr/chinese-national-residing-california-arrested-theft-artificial-intelligence-related-trade) [https://perma.cc/EVU5-5ASX].

273 Hrdy, *supra* note 251, at 113.(introducing that like model extraction attacks, knowledge distillation uses large volumes of a teacher model's outputs to train new model that approximates the original's behavior).

274 Karim, *supra* note 268.

275 Hrdy, *supra* note 251, at 113.

276 Hrdy, *supra* note 251, at 85–86; *see also* Henderson & Lemley, *supra* note 90, at 1360–61 (claiming that whether using one model's outputs to train a competitor constitutes traditional reverse engineering is unclear; even if it did, courts are divided on enforcing contractual bans, and trade secret law expressly permits reverse engineering).

Reverse engineering is conventionally defined as "starting with the known product and working backward to find the method by which it was developed."[277] In the context of knowledge distillation, although the student model learns from the teacher model's outputs, the resulting student and teacher models typically demonstrate markedly different inference pathways.[278] Thus, the purpose and methodology of knowledge distillation are fundamentally distinct from those found in reverse engineering. Indeed, the similarity between a product created through reversed engineering and its original counterpart is often substantially greater than that which exists between a teacher model and its distillation-driven student model.

Courts have consistently recognized reverse engineering as a lawful and legitimate method for analyzing publicly available products or information, enabling their recreation or improvement.[279] The DTSA also explicitly states that reverse engineering does not constitute an "improper means" of acquiring a trade secret.[280] This exclusion reflects the broadly accepted principle that reverse engineering should be permitted to promote innovation and competition.[281]

Both reverse engineering and knowledge distillation seek to understand and emulate the functionality of an original model or product. However, due to the inherently opaque "black-box" character and the dynamic complexity of modern AI systems, knowledge distillation presents formidable challenges as compared to traditional reverse engineering. In most cases, a student model develops its own inference patterns from only a limited understanding of the teacher model's reasoning, which itself is informed by undisclosed training data and intricate architectures. In other words, knowledge distillation does not attempt to reconstruct the teacher model in its entirety, and the outcome is almost always less precise than a product produced by conventional reverse engineering. Therefore, if reverse engineering is not an "improper means," there is a stronger reason to exclude knowledge distillation from the scope of "improper means."

Given that reverse engineering is expressly recognized as a proper means of acquiring trade secrets under U.S. law, there is even greater reason to exclude knowledge distillation from the definition of "improper means." The process is further attenuated and less exact; the resulting student models do not faithfully replicate their teacher models but instead internalize general patterns and develop independent reasoning mechanisms. Therefore, we argue that most forms of knowledge distillation, as presently practiced in the field of AI, do not amount to trade secret misappropriation.

---

[277] UNIF. TRADE SECRETS ACT § 1 cmt. 1 (UNIF. L. COMM'N) (amended 1985).

[278] Prajjwal Bhattarai et al., *Knowledge Distillation through Geometry-Aware Representational Alignment*, CORN. U., Sep. 27, 2025, at 1, arXiv:2509.25253v1 (concluding that whereas traditional distillation aligns output distribution via probabilistic divergence, feature-based approaches minimize Euclidean variants between hidden representation to make the student mimic the teacher's feature space).

[279] *See, e.g.*, Pamela Samuelson & Suzanne Scotchmer, *The Law and Economics of Reverse Engineering*, 111 YALE L.J. 1575, 1582 (2002).

[280] *See* 18 U.S.C. § 1839(6)(B).

[281] *See, e.g.*, Kewanee Oil Co. v. Bircron Corp., 416 U.S. 470, 476 (1974); Lena Chan, *Note: The Weaponization of Trade Secret Law*, 124 COLUM. L. REV. 703, 713–14 (2024); *see also* Chi. Lock Co. v. Fanberg, 676 F.2d 400, 405 (9th Cir. 1981) (noting the reverse engineering exception prevents trade secrets from being "convert[ed] . . . into a state-conferred monopoly akin to the absolute protection that a federal patent affords").

### 2. Copyright

Copyright infringement may appear to be a potential cause of action if the developer of a teacher model wishes to assert a legal right over the output produced by that model.[282] However, the question of whether AI-generated content is copyrightable has been the subject of extensive debate in various jurisdictions in recent years.[283] In the United States, the law has become increasing clear: AI-generated output is not eligible for copyright protection.

The U.S. Copyright Office unequivocally stated in a report published in January 2025 that materials created solely by inputting prompts into an AI system lacks requisite human expressive elements and, therefore, cannot be copyrighted.[284] This position is consistent with the ruling of the U.S. Court of Appeals for the District of Columbia Circuit's in *Thaler v. Perlmutter*, which held that works generated autonomously by AI systems are not copyrightable.[285]

Consequently, since AI-generated output does not enjoy copyright protection, the developer of a teacher model cannot allege copyright infringement against the developer of a student model on the basis of unauthorized use of output generated by the teacher model.[286] The absence of legal protection for such outputs further diminishes the likelihood of copyright serving as a viable remedy in disputes arising from knowledge distillation.

### 3. Terms of Use

AI development companies whose models may serve as teacher models in knowledge distillation may seek to protect their models through contractual provisions.[287] This strategy parallels to the practice of companies using contracts to prohibit reverse engineering, a tactic adopted partly in response to the challenges of pursuing trade secret misappropriation claims under the law.[288] Yet, scholars and courts remained divided on the legality of restricting reverse engineering through contracts,[289] and the question is even more controversial when it comes to prohibiting knowledge distillation by contractual arrangements.

---

[282] Ferguson, Liu & Han, *supra* note 62.

[283] *See, e.g.*, Jyh-An Lee, *Computer-Generated Works Under the CDPA 1988*, *in* ARTIFICIAL INTELLIGENCE AND INTELLECTUAL PROPERTY 177 (Jyh-An Lee, Reto M. Hilty & Kung-Chung Liu eds. 2021); Rita Matulionyte & Jyh-An Lee, *Copyright in AI-Generated Works: Lessons from Recent Developments in Patent Law*, 19 SCRIPTED: J. L. TECH. & SOC'Y 5, 12–22 (2022).

[284] U.S. COPYRIGHT OFFICE, COPYRIGHT AND ARTIFICIAL INTELLIGENCE: PART 2 – COPYRIGHTABILITY 18 (2025).

[285] Thaler v. Perlmutter, 130 F.4th 1039, 1041 (D.C. Cir. 2025) (affirming the denial of a copyright application because A.I. "cannot be the recognized author of a copyrighted work because the Copyright Act of 1976 requires all eligible work to be authored in the first instance by a human being.").

[286] Philipp, *supra* note 90, at 68; Karim, *supra* note 268.

Unsurprisingly, DeepSeek would have violated OpenAI's Terms of Use if it employed ChatGPT-generated outputs to train its own models.[290] Melissa Holyoak, a Commissioner of the Federal Trade Commission ("FTC") echoed this allegation, claiming that DeepSeek engaged in unfair competition by using ChatGPT-generated outputs and thereby violating OpenAI's policies.[291] Such allegations rely on contractual provisions—including anti-reversing engineering and anti-compete clauses—commonly adopted by leading LLM developers such as OpenAI, Google, and Meta to prevent users from reverse engineering their models or using model outputs to compete against them.[292]

Despite the prevalence of these restrictions, experts hold diverse views on their enforceability. Professor Camilla A Hrdy has suggested that courts are more likely to enforce the non-compete clauses found in business-to-business agreements than those imposed upon individual users.[293] She contends, moreover, that even if these clauses are ultimately found unenforceable, their mere existence can exert a "chilling effect" by deterring potential rivals through the threat of litigation.[294]

---

[287] Philipp, *supra* note 90, at 77–78.

[288] *See, e.g.*, Mark A. Lemley, *The Surprising Virtues of Treating Trade Secrets as IP Rights*, 61 STAN. L. REV. 311, 350–51 (2008).

[289] *See, e.g.*, Hrdy, *supra* note 251, at 80 (noting that under trade secret law, reverse engineering is permissible; however, companies frequently attempt to restrict it through contractual agreements, and courts commonly uphold these provisions.); *id.* at 361, (arguing against "parties... opting out of particular rules of trade secret law, at least to the extent they rely on trade secret rather than contract remedies"); Samuelson & Scotchmer, *supra* note 272, at 1626–27, 1660 (noting that while software licenses typically forbid reverse engineering, the enforceability of such contractual restrictions remains an unsettled legal issue, with courts in the United States and internationally divided on the matter).

[290] *See, e.g.*, Ariel Zilber & Thomas Barrabi, *OpenAI Says It Has Proof DeepSeek Used Its Technology to Develop Its AI Model*, N.Y. POST (Jan. 29, 2025, at 12:36 ET), https://nypost.com/2025/01/29/business/openai-says-it-has-proof-deepseek-used-its-technology-to-develop-ai-model [https://perma.cc/Z49N-TASW]; Criddle & Olcott, *supra* note 31.

[291] Bryan Koenig, *FTC's Holyoak Has Her Eyes on DeepSeek*, LAW360 (Feb. 21, 2025), https://www.law360.com/articles/2300869/ftc-s-holyoak-has-her-eyes-on-deepseek [https://perma.cc/LRG3-LFFK].

[292] *See, e.g.*, OpenAI, *Term of Use*, https://openai.com/policies/row-terms-of-use [https://perma.cc/5Z6C-8MRZ] (last visited Mar. 1, 2025, at 14:46 CT) ("you many not . . . [a]ttempt to or assist anyone to reverse engineer, decompile or discover the source code or underlying components of our Services, including our models, algorithms, or systems (except to the extent this restriction is prohibited by applicable law) . . . [a]utomatically or programmatically extract data or Output . . . [u]se Output to develop models that compete with OpenAI); Google, *Gemini API Additional Terms of Service*, https://ai.google.dev/gemini-api/terms [https://perma.cc/YQ4Y-EKSK] (last visited Mar. 1, 2025, at 14:47 CT) ("You may not use the Services to develop models that compete with the Services . . . [y]ou also may not attempt to reverse engineer, extract or replicate any component of the Services, including the underlying data or models (e.g., parameter weights)); Meta, *Term of Use*, https://ai.meta.com/llama/license [https://perma.cc/2Z5M-54X9] (last visited Mar. 1, 2025, at 14:48 CT) ("You will not use the Llama Materials or any output or results of the Llama Materials to improve any other large language model . . . .").

By contrast, Professors Peter Henderson and Mark A. Lemley, drawing on the Second Circuit's decision in *ML Genius Holdings LLC v. Google LLC*, have argued that such provisions would encounter significant challenges in most circuits, if used to restrict knowledge distillation.[295] Legally, they explain, the absent copyright protection for teacher model outputs means that claims based in state contract law are likely preempted by copyright law.[296] From a policy perspective, they further argue that companies such as OpenAI—who themselves have scraped content from billions of web pages in violation of site terms of use, invoking "fair use" in their defense—cannot fairly claim the right to contractually prohibit similar behavior by their competitors.[297]

The parallel Henderson and Lemley draw from *Genius* is particularly compelling. In *Genius*, the company sought to enforce anti-scraping clauses against Google despite not owning the rights to the data in question.[298] Likewise, in the context of knowledge distillation, it is unlikely that the developer of a teacher model can actually own the copyright of its outputs. These circumstances make it exceedingly difficult for OpenAI to enforce anti-competition provisions against DeepSeek, not to mention the challenge of proving the exact source and nature of DeepSeek's knowledge distillation.

Consequently, although contractual restrictions are a common mechanism for asserting control over AI models and their outputs, their legal enforceability—particularly in the nuanced context of knowledge distillation—remains deeply uncertain, both in principle and in practice.

## IV. CONCLUSION

The disruptive impact of the DeepSeek model is fundamentally anchored in the advanced technologies it brings together. By refining reinforcement learning techniques, DeepSeek dramatically reduces its reliance on computational power and extensive training data. DeepSeek's efficiency is further enhanced by its use of the MoE architecture, which combines fine-grained expert segmentation with a shared expert isolation mechanism, allowing the model to allocate resources judiciously. The deployment of knowledge distillation techniques enables DeepSeek to achieve high performance while substantially lowering training costs. Equally significant is DeepSeek's open-source ethos, which invites diverse forms of customization and enhances transparency, fostering continuous optimization and helping to mitigate bias. However, this openness also introduces notable safety concerns, including vulnerabilities to jailbreaking and other forms of cyberattack.

---

[293] Hrdy, *supra* note 251, at 80.

[294] *Id.*

[295] Henderson & Lemley, *supra* note 90, at 1367–68.

[296] *Id.* at 1368.

[297] *Id.* at 1369; *see also* Criddle & Olcott, *supra* note 31 ("OpenAI is battling allegations of its own copyright infringement from newspapers and content creators, including lawsuits from The New York Times and prominent authors, who accuse the company of training its models on their articles and books without permission."); *But cf.* Frank Pasquale, Thomas W. Malone, & Andrew Ting, *Copyright, Learnright, and Fair Use: Rethinking Compensation for AI Model Training*, 23 NW. J. TECH. & INTELL. PROP. 205, 214–15 (2025) (arguing that "when advocates for U.S. AI firms accused the Chinese firm DeepSeek of stealing their intellectual property, they were widely ridiculed for hypocrisy.").

[298] ML Genius Holdings LLC v. Google LLC, No. 20-3113, 2022 WL 710744, at *3 (2d Cir. Mar. 10, 2022).

DeepSeek's success has generated contentious legal and policy debates, magnified by ongoing geopolitical tensions between the United States and China. Its breakthrough has raised pointed questions about the efficacy of U.S. export controls on AI hardware—such as chips and semiconductors—which are now subject to continual recalibration, informed by global power dynamics, industry lobbying, and rapid technological evolution. While worries surrounding information security are not unique to DeepSeek, its Chinese origin has heightened scrutiny and led several governments to impose restrictions or outright bans.

The model has also faced criticism for the political and ideological bias reflected in its outputs—a concern that is, however, not isolated to DeepSeek, but shared among various LLMs. Given DeepSeek's strengths in areas such as Chinese natural language processing, clinical decision-making, and competitively priced APIs, such bias may have a limited impact on adoption in certain markets. Nevertheless, expanding its global reach will require a deliberate effort to address and rectify these issues.

Finally, DeepSeek's cost efficiency is intrinsically linked to its innovative use of knowledge distillation. Whether or not DeepSeek has distilled proprietary models, the foundational nature of this technology poses significant challenges for creators of proprietary models in claiming their rights under IP or contract law.